\documentclass[aps,pre,floatfix,nofootinbib,showkeys]{revtex4}
\usepackage{enumerate}
\usepackage{subfigure}
\usepackage{epic}\usepackage{eepic}
\usepackage{bm}
\usepackage{color}
\usepackage{xcolor}
\usepackage{soul}

\usepackage[dvips]{epsfig}
\usepackage{graphicx} \usepackage{amsmath} \usepackage{amssymb}
\newcommand{\comment}[1]{}

\newcommand{\ve}{\varepsilon}

\newcommand{\A}{{\cal A}}
\newcommand{\B}{{\cal A}'}

\newcommand{\be}{\begin{eqnarray}}
\newcommand{\ee}{\end{eqnarray}}

\renewcommand{\d}{{\rm d}}
\newcommand{\no}{\label}

\begin{document} 

\title{
Adaptive Decision Making via Entropy Minimization}

\author{ Armen E. Allahverdyan$^{1)}$, Aram Galstyan$^{2)}$, Ali E. Abbas$^{3)}$, and Zbigniew R. Struzik$^{4,5)}$ 
\footnote{armen.allahverdyan@gmail.com, galstyan@isi.edu, aliabbas@price.usc.edu, z.r.struzik@p.u-tokyo.ac.jp\\
Published in International Journal of Approximate Reasoning, {\bf 103}, 270-287 (2018). }}

\address{ 
$^{1)}$Yerevan Physics Institute, Alikhanian Brothers Street 2,  Yerevan 375036, Armenia\\
$^{2)}$USC Information Sciences Institute, 4676 Admiralty Way, Marina del Rey, CA 90292, USA\\
$^{3)}$Industrial and Systems Engineering, University of Southern California, Los Angeles, USA,\\
$^{4)}$Advanced Center for Computing and Communication, RIKEN Institute, 2-1 Hirosawa, Wako-shi, 351-0198, Japan,\\ 
$^{5)}$Graduate School of Education, The University of Tokyo, 7-3-1 Hongo, Bunkyo-ku, Tokyo, 113-0033, Japan. 
}
 
\begin{abstract} 
 
An agent choosing between various actions tends to take the one with the
lowest cost. But this choice is arguably too rigid (not adaptive) to be
useful in complex situations, e.g., where exploration-exploitation
trade-off is relevant in creative task solving or when stated
preferences differ from revealed ones.  Here we study an agent who is
willing to sacrifice a fixed amount of expected utility for adaptation.
How can/ought our agent choose an optimal (in a technical sense) mixed
action? We explore consequences of making this choice via entropy
minimization, which is argued to be a specific example of risk-aversion.
This recovers the $\epsilon$-greedy probabilities known in reinforcement
learning. We show that the entropy minimization leads to rudimentary
forms of intelligent behavior: {\it (i)} the agent assigns a
non-negligible probability to costly events; but {\it (ii)} chooses with
a sizable probability the action related to less cost (lesser of two
evils) when confronted with two actions with comparable costs; {\it
(iii)} the agent is subject to effects similar to cognitive dissonance
and frustration.  Neither of these features are shown by entropy
maximization. 

\end{abstract}
 

\maketitle 

\section{Introduction}

Consider an agent who has to choose between a number of different
actions $\A_1,...,\A_n$. Before taking action, consequences 
of each action $\A_k$ are subjectively estimated to have the cost
$\varepsilon_k$ (or the utility $u_k=-\varepsilon_k$). The basic tenet
of decision theory is that the agent ought to choose the action that
minimizes the cost (or maximizes the utility)~\cite{luce}. In terms of
probabilities $p_k$ for various actions ($p_k\geq 0$, $\sum_{k=1}^n
p_k=1$), this amounts to taking the action $\ell$ related to the least
cost (if it exists and is available):
\begin{eqnarray}
  \label{eq:1}
  p_{\ell}=1, \qquad p_{k\not=\ell}=0, \qquad
\ve_\ell<\ve_{k\not=\ell},\qquad k=1,...,n.  \end{eqnarray} 
There are, however, situations where $\varepsilon_k$ may change after
actions are taken, also as a result of those actions \footnote{See
section \ref{stato} for more details.  We stress that we do not mean the
delayed reward situation, where the utility is constant, but is
discounted by some known factor, because the action is performed now,
while its reward will come in future.}. Here is an example that points
against choosing (\ref{eq:1}), and illustrates our problem. You got 100
eggs and several baskets, which seem to have different durabilities (i.e.
utilities). The probability with which an action, i.e. a basket, is taken 
refers to the fraction of eggs in it.
Even if the most durable basket may appear to support all the
eggs, it is not wise to put everything in one 
basket. First the durability of a basket can change due to very
eggs inside of it. Second, the durability can change unexpectedly due to
hindrances. Third, you loose the possibility to explore other baskets
that may turn out to be more durable than you thought. Let us now
mention several more, broadly defined situations, where ``putting all
eggs into one basket'' is not good. 

-- In reinforcement learning the preliminary costs $\ve_k$ do change due
to the actions taken \cite{barto}. Even if these changes are assumed to
be predictable, the agent still needs to make several actions before
enough experience is accumulated. 

-- The exploration-exploitation dilemma is known in adaptive
(biological, organizational, social) systems; see
\cite{exp_exp,exp_exp_1} for reviews. Exploring possibilities that seem
inferior from a local viewpoint may provide advantages in the long run.
Exploitation (in the narrow sense) makes the choice that does seem
optimal at the moment of choice. Broader exploitation scenarios do
account for adaptivity, but still concentrate on the most useful
possibilities~\cite{exp_exp_1}. 

-- In creative problem solving there are conceptually simple tasks which
are nevertheless not easy to solve in practice because solving them via
the least cost (implied by the statement of the problem and/or the
previous experience of the solver) is a dead
end~\cite{handbook,maier,luchins}. This {\it Einstellung} effect is one
of the main hindrances of human
creativity~\cite{handbook,maier,luchins}. Creative tasks can be solved
only if (subjectively) less probable ways are looked
at~\cite{handbook,maier}. 

\comment{Another limitation of (\ref{eq:1}) is that it assumes that
$\ve_\ell={\rm min}_k[\varepsilon_k]$ is readily available, which may not be the
case in practice, since finding this minimum may be related to
complexity cost (if $n$ is large).}

How to assign prior probabilities to avoid the strictly deterministic
(\ref{eq:1})?  Such probabilities should hold a natural constraint that
actions related to higher cost are getting smaller probabilities.  Two
ad hoc solutions are especially simple: one can take into account only
the second-best action, or take all non-best actions with the same
(small) probability. In reinforcement learning the latter prior
probability is known as the $\epsilon$-greedy~\cite{barto}. It is
preferable to have a regular method of choosing non-deterministic
probabilities, which will reflect people's attitudes towards the
decision making in an uncertain situation, and which will include the
above ad hoc solutions as particular cases. 

% (We work with costs or minus utilities to make stronger potential
% analogies with statistical physics, where the cost is akin to energy.)

% As explained below, yet another reason for employing probabilistic
% decisions corresponds with choices between lotteries.

Here we explore the possibility of defining the prior probabilities via
risk minimization (or maximization); see~\cite{lola,lopes} for reviews
on the notion of risk and its various interpretations. We assume that
the agent first decides how much average utility $E-{\rm min}_k[\ve_k]$
he invests into exploration by going into nonoptimal|in the sense of
not holding (\ref{eq:1})|behavior. We employ the notion of risk in a
specific context, namely  when comparing the behavior of agents
having the same utilities for various actions and the same value of $E$.
We argue below that maximizing (minimizing) risk in this specific
situation can be done via maximizing (minimizing) the entropy
$-\sum_{k=1}^np_k\ln p_k$. People demonstrate both risk minimization
(aversion) and maximization (seeking)~\cite{tver,baron}, though the risk
in those situations is a less specific (and more difficult to describe)
notion---first because it involves agents having different utilities for
same actions, and second because it involves a difference between the
monetary value (gain or loss) and its utility. 

Our results show that there are important behavioral differences
between entropy-minimizing and entropy-maximizing agents.  They are seen
for at least three different actions (and the same $E$). The
entropy-minimizing agent implements risk-aversion by weighting the
least-cost action more, but he also assigns a non-negligible probability for
the high-cost action---whereas the entropy-maximizing agent ignores it.
The extent to which the high-cost action is accounted for by the
entropy-minimizing agent depends on the amount of utility invested into
exploration: investing more utility leads to assigning less probability.
As we argue below, this closely relates with the notion of cognitive
dissonance \cite{disso,aronson}. Another feature is frustration: due to
competing local minima of entropy, the entropy-minimizing agent can
abruptly change the action probabilities as a result of a small change
of $E$. Also, when confronted with two actions with different, but
comparable costs, the entropy-minimizing agent tends to select the one
with a smaller cost (chooses the lesser of two evils), while the
entropy-maximizing agent simply does not distinguish between them. The
important point is that for a risk-minimizing agent (which does a
constrained minimization of a concave function in a convex domain)
choosing the probabilities of actions means selecting between several
local minima. In contrast, the risk-seeking agent always has a unique
and well-defined probabilistic solution that results from minimizing a
convex function~\cite{rock}. We relate the above features of the
entropy-minimizing agent with a rudimentary form of intelligence (see
Section \ref{summa}). 

The remainder of this paper is organized as follows. Section \ref{stato}
explains the statement of our problem. Section \ref{rii} discusses
stochastic dominance, risk, majorization and its relation with entropy.
In particular, Section \ref{maxomino} provides general remarks and
references on entropy optimization. The reader who agrees from the
outset with entropy as a measure of risk (and uncertainty) can consult
Sections \ref{stato} and \ref{rii} very briefly. The next two sections,
\ref{min} and \ref{maxent}, present details of (resp.) entropy
minimization and maximization.  The latter may seem to be standard, but
Section \ref{maxent} still contains salient points that are frequently
overlooked. Section \ref{compa} compares entropy maximization and
minimization scenarios from the viewpoint of the agent's behavior. We
summarize in Section VII. 

% \footnote{We employ the standard terminology, where the decision under
%   known probabilities is called risk, partially known probabilities
%   indicate on uncertainty, while completely unknown (but non-zero)
%   probabilities mean ignorance. }

\section{Statement of the problem}
\no{stato}

\subsection{Costs and constraints}
\no{coco}

Let us explain the above problem. An agent faces different
actions $\{\A_k\}_{k=1}^n$ with (resp.) costs $\{\ve_k\}_{k=1}^n$.
These costs are 
subjective estimates of future consequences of actions 
made by the agent before deciding on those actions. After taking
several actions, the agent can change his estimates also as a result of
the actions taken. However, before taking actions he does not know in
which specific way the costs will change. In such an agnostic situation,
the agent faces two normative demands---he should behave according to
$\{\ve_k\}_{k=1}^n$, but he also should also explore all actions. Hence
he decides to act via probabilities and implements two constraints.
First, he decides to invest in the exploration of  the average utility
$E-{\rm min}_k[\varepsilon_k]$, where
\begin{eqnarray}
  \label{eq:4}
  E\equiv\sum_{k=1}^n p_k\, \varepsilon_k , \qquad p_k\geq 0, \qquad \sum_{k=1}^n p_k=1,
\no{osh}
\end{eqnarray} 
and where $p_k$ are probabilities to be chosen.  Within this formulation
of the problem, we can regard $E$ to be under control of the agent. 

Second, actions related to a lesser cost get higher probability:
\begin{eqnarray}
  \label{eq:5}
  (p_k-p_l)(\varepsilon_k-\varepsilon_l)\leq 0.
\end{eqnarray}
Recall that costs are negative utilities. We work with costs, since this
makes explicit the analogy with statistical physics, where the cost
corresponds to energy, and the natural tendency is to minimize the
energy.\footnote{Note that (\ref{eq:5}) is the standard condition of
physical stability~\cite{lenard} Relations between statistical physics
and decision theory are mentioned in~\cite{yuko}.  Ref.~\cite{abbas}
discusses the maximum entropy method for elicitation of costs
(utilities) from an incomplete data. }

We order the costs and probabilities as 
\begin{eqnarray}
  \label{eq:08}
  \varepsilon_1< \varepsilon_2< ....< \varepsilon_n, \\ 
\label{borik}
p_1\geq p_2\geq ...\geq p_n.  \end{eqnarray} Note that the inequalities in (\ref{eq:08})
between costs are strict, since within the present study coinciding costs mean coinciding
actions. Hence once the actions are different, so should be the costs. 
The reason for insisting on non-equal costs in (\ref{eq:08}) will be seen below. 

\subsection{Stated versus revealed preferences}
\no{marlo}

We can look at this problem from a different angle. In human
decision-making, it is known that preferences with respect to different
actions can be determined in laboratory conditions (e.g., via surveys).
This then defines their costs (negative utilities) (\ref{eq:08}).
However, in reality (e.g., in complex conditions of a market) people
generally do not strictly follow these preferences due to lack of
cognitive abilities, insufficient attention, lack of a decision time,
the need of an exploration behavior etc.. The difference between
experimental and real-world choices is known as the problem of stated
versus revealed preferences; see \cite{marley} for a recent review.  It
is assumed in the psychology of choice that such cases can be described
by assigning probabilities to each costs \cite{marley}. Now, the above
formulation does correspond to this situation, where (\ref{borik}) is a
natural constraint on probabilities, and where $E-{\rm
min}_k[\varepsilon_k]>0$ is not simply chosen by the agent, but is also
due to objective problems described above. For $E-{\rm
min}_k[\varepsilon_k]\to 0$, the agent is back to choosing the
least-cost solution only. 

\subsection{Differences with respect to other models of decision theory}

Above we described an agent who is uncertain about future consequences
of his actions. Decision theory has models for that situation. The
classic model (by Savage and others) \cite{luce,jeffrey,baron} assumes
that at the moment of action-taking there is an uncertain state of
nature (environment) ${\cal S}_\alpha$ to be realized from $\{{\cal
S}_\alpha\}_{\alpha=1}^m$ with probabilities
$\{\pi_\alpha\}_{\alpha=1}^m$, which can be the agent's subjective degrees
of belief. These are called states of nature, since their future
realization is independent from the action taken, but an action $\A_{i}$
in a state ${\cal S}_\alpha$ leads to consequences with costs
$c_{i\alpha}$ \cite{luce,jeffrey,baron}. The agent does know
$\{\pi_\alpha\}_{\alpha=1}^m$ and
$\{c_{i\alpha}\}_{i=1\,\alpha=1}^{n\quad m}$. Now one criterion for looking
for the best action is to choose $i$ such that the expected cost
$\sum_{\alpha=1}^m \pi_\alpha c_{i\alpha}$ is minimized over $i$ (i.e.,
the expected utility is maximized) \cite{luce,jeffrey,baron}. 

The classic model has limitations---e.g., that environmental states
$\{{\cal S}_\alpha\}_{\alpha=1}^m$ are realized independently from
actions. There are many cases where this simplistic assumption does not
hold. Causal decision theory \cite{giba,joyce,skyrms,rati,shafer}
attempts to remedy this problem, but creates its own issues---e.g.,
replacing $\pi_\alpha$ by the conditional probability $\pi_{\alpha|i}$
of the state ${\cal S}_\alpha$ given the action $\A_i$ leads to
paradoxical (if not unacceptable) conclusions nicknamed ``voodoo''
decisions \cite{skyrms}. In response, the expected utility principle was
changed to the concept of ratifiability \cite{jeffrey,rati}, but this
proposal has its own problems \cite{rati}, e.g.. it demands an unusual
environment.\footnote{Ratifiability advises to take action $\A_i$ if for
all $j$ \cite{jeffrey,skyrms,rati}: $C(\A_j|\A_i)\geq C(\A_i|\A_i)$,
where $C(\A_j|\A_i) \equiv\sum_\alpha \pi_{\alpha|i} c_{j\alpha}$.
(Minimization of the expected cost reads in this notation
$C(\A_j|\A_j)\geq C(\A_i|\A_i)$.) This assumes an unusual environment:
$C(\A_j|\A_i)$ describes an agent who commits to act $\A_i$,
conditionally constrains the environment by this commitment, but still
keeps the freedom of acting $\A_j$. The ratifiability then means
consistency between the commitment and the actual action.}  Another
response was to change the conditional probability $\pi_{\alpha|i}$ to
the probability of conditional $\pi_{\alpha \Leftarrow i}$, which, in 
contrast to $\pi_{\alpha|i}$, is supposed to describe the causal
influence of the action $\A_i$ on ${\cal S}_\alpha$ \cite{joyce,giba}.
But $\pi_{\alpha \Leftarrow i}$ does not have a general definition
\cite{giba}, while some of its particular definitions are flawed
\cite{shafer}.  Recent reviews show that all non-classical models of 
causal decision theory are problematic in one way or another
\cite{rati,giba,shafer}. 

Our statement of the problem refers to an agnostic agent that (as yet) does
not know about possible consequences of his actions in a complex and
uncertain world. Even if he is going to learn about them, he still
needs to perform several actions with certain prior probabilities. We do
not assume any specific mechanism by which uncertain states of the world are
realized and changed in response to actions. Hence, our model can be at
best descriptional, but it starts from two normative demands: behave
according to your own estimated costs (\ref{eq:08}) and explore all
actions. 

\section{Risk and entropy}
\label{rii}

\subsection{Stochastic dominance, risk and majorization}

Together with probabilities $p_1,..., p_n$ in (\ref{eq:4}--\ref{borik}), we
define another candidate probability with the same costs (\ref{eq:08})
and holding the same constraints (\ref{eq:4}, \ref{eq:5}) with the same value of $E$:
\be
\label{mavrik}
q_1\geq q_2\geq ...\geq q_n\geq 0, \quad \sum_{k=1}^n q_k=1, \quad 
  E=\sum_{k=1}^n q_k\, \varepsilon_k .
\ee
We ask for criteria that should make $\{p_k\}^n_{k=1}$ more preferable
than $\{q_k\}^n_{k=1}$. The well-known criterion of choosing for a
smaller average cost obviously does not apply here. However, one can ask
whether $\{p_k\}^n_{k=1}$ is less risky than $\{q_k\}^n_{k=1}$.  

We emphasize that risk is generally a complex notion that raises
conceptual and practical (resp. normative and descriptional) issues; see
\cite{lola,lopes,pope,baron,aumann,leshno,tver,chisar,ahmad,chin1,chin2,ng}
for reviews from various viewpoints. Our application of risk will be to
a large extent free of those issues. There are two reasons for this.
First, we do not focus on the difference between the commodity and its utility (cost);
hence we are not concerned with the problem of convex/concave utility functions
\cite{baron,tver}.
Second, we always compare between lotteries (i.e., probabilities and
their outcome costs) that are ordered in the same way
[cf.~(\ref{mavrik}, \ref{borik})] and refer to the same costs
(\ref{eq:08}). Then a stringent definition of risk goes via the stochastic dominance
condition\footnote{This is the first-order stochastic dominance
condition~\cite{luce,haim}. Second and higher-order conditions refer to
the situation, where there are two different concepts related to the
cost (or negative utility): money and its proper utility. We
do not employ them here. }; see~\cite{luce,haim} for reviews. A risk-averse
agent can start with the largest possible cost $\varepsilon_n$ and asks
for its $p$-probability $p_n$ to be not larger than the $q$-probability
$q_n$. Next, the $p$-probability $p_{n}+p_{n-1}$ of the cost to be
larger or equal to $\varepsilon_{n-1}$ is compared with the
corresponding $q$-probability $q_{n}+q_{n-1}$ etc. We end up with $n-1$
conditions $\sum_{i=n}^k p_{i}\leq \sum_{i=n}^k q_{i}$, or
\be
\sum_{i=1}^k p_{i}\geq \sum_{i=1}^k q_{i}, \qquad k=1,...,n-1,
\label{major}
\ee
so that if one of them holds strictly, then $\{p_k\}^n_{k=1}$ is less
risky than $\{q_k\}^n_{k=1}$. We emphasize that the same $n-1$ conditions
(\ref{major}) emerge with an alternative definition of risk, when the
agent maximizes the probability of the best outcome $\varepsilon_1$
etc. So within the present definition it does not matter whether a 
risk-averse agent selects for higher probabilities of lower-cost options, 
or for lower probabilities of higher-cost actions. 

Due to (\ref{eq:08}, \ref{borik}, \ref{mavrik}), the stochastic
dominance condition (\ref{major}) coincides with
the majorization condition~\cite{major}.\footnote{We stress the
difference between majorization and stochastic dominance: in the latter
the probabilities are not ordered, only the costs are ordered as in
(\ref{eq:08}). For majorization the probabilities should be ordered as
in (\ref{borik}). Given this difference in orderings, both majorization
and stochastic dominance are defined by (\ref{major}). Generally, the
stochastic dominance does differ from majorization~\cite{luce,haim}. }
\be
\no{majo}
\{p_k\}_{k=1}^n\succ \{q_k\}_{k=1}^n. 
\ee
In particular, for any $\{p_k\}_{k=1}^n$ we get 
$\{p_k\}_{k=1}^n\succ \{q_{k}=\frac{1}{n}\}_{k=1}^n$ (once
$q_k$ holds the first condition in (\ref{mavrik})), while
$(1,0,..,0)\succ \{p_k\}_{k=1}^n$. Thus, in the
present situation we have a natural correspondence between risk and
majorization. 

\subsection{Measure of risk and Schur-concave functions}

According to (\ref{major}), we have $n-1$ different conditions; i.e., it
is not a single number as a measure for risk.\footnote{The literature on
mathematical economics suggests certain single-number measures of risk;
see e.g.,~\cite{aumann}.  They do not apply for our situation since
they demand that $E\leq 0$ (the average gain), but there are
certainly some indices $i$ for which $\varepsilon_i>0$. This condition
is too restrictive for us. Ref.~\cite{leshno} presents another
interesting attempt to go beyond the stochastic dominance conditions.}
Hence even if (\ref{major}) holds, i.e.,  $\{q_k\}_{k=1}^n$ is more risky
that $\{p_k\}_{k=1}^n$, we do not have any obvious way of quantifying
this relation via a single number or determine its approximate validity. 

A more serious drawback of (\ref{major}) is that for many interesting
cases the risk is simply not defined, since neither
$\{p_k\}_{k=1}^n\succ \{q_k\}_{k=1}^n$ holds nor $\{q_k\}_{k=1}^n\succ
\{p_k\}_{k=1}^n $:
\be
\{p_k\}_{k=1}^n\not\succ \{q_k\}_{k=1}^n\quad {\rm and}\quad
\{q_k\}_{k=1}^n\not\succ \{p_k\}_{k=1}^n. 
\label{tatar}
\ee
For $n=3$, (\ref{tatar}) is equivalent to
\be
\label{abver}
(p_1-q_1)(p_3-q_3)>0. 
\ee

These drawbacks motivate us to take (\ref{major}) as a sufficient
condition for risk, but still look for a single-function measure of risk
${S}[p_1,...,p_n]$ such that if (\ref{major}) holds, then
\be
\no{stary}
{S}[p_1,...,p_n]\leq {S}[q_1,...,q_n].
\ee
While ${S}[p_1,...,p_n]$ is initially defined
for ordered probabilities (\ref{borik}), it is natural to extend this
definition by demanding that ${S}[p_1,...,p_n]$ is invariant with
respect to any permutation of its $n$ arguments. Functions 
that hold these two conditions|permutation invariance and 
that (\ref{majo}) implies (\ref{stary})---are called Schur-concave\footnote{Multiplying by minus one makes such a
function Schur convex. } \cite{major}. For a differentiable function to
be Schur-concave it is necessary and sufficient that \cite{major}
\be
(p_i-p_j)\left(\frac{\partial S}{\partial p_i}-
\frac{\partial S}{\partial p_j}\right)\leq 0.
\label{schur}
\ee
Naturally, the choice of a Schur-concave function as a measure of risk
is not unique; hence certain additional conditions are needed to fix it.
Below we discuss one set of such conditions that allows us to fix entropy
as a measure of risk \cite{leod}. But before doing so, let us stress that
a candidate Schur-concave function will in fact define the risk for
$\{p_k\}_{k=1}^n$ and $\{q_k\}_{k=1}^n$ that hold (\ref{tatar}).  For
illustration, let us take the simplest non-trivial case of $n=3$. Now
(\ref{abver}) shows that the risk is to be defined between probabilities
that have a conflict between giving more probability to the least-cost action
$\A_1$ or the less probability to the worst-cost action $\A_3$; i.e., we face a
conflict between two different aspects of risk-taking, to be resolved
via a reasonable choice of the Schur-concave function ${S}[p_1,...,p_n]$.
Note in this context that according to (\ref{schur}) there cannot be a
differentiable ${S}[p_1,...,p_n]$ for which, e.g., ${S}[p_1,...,p_n]
<{S}[q_1,...,q_n]$ whenever $p_3={\rm min}[p_1, p_2, p_3]<q_3={\rm
min}[q_1, q_2, q_3]$. 

\subsection{ Entropy}

Following \cite{leod,presse}, we now outline several additional
conditions that lead to choosing entropy among other Schur-concave
functions. 

\begin{enumerate}[{(A)}]

\item \label{A}
$S[p_1,...,p_n]$ is additive with respect to the index $k$, i.e.,
\be
\no{adon}
S[p_1,...,p_n]=\sum_{k=1}^n\psi(p_k), 
\ee
with a smooth function $\psi(x)$; cf.~(\ref{conv}). Condition
(\ref{adon}) can be motivated from demanding that if we compare $(p_1,...,p_n)$ and  $(q_1,...,q_n)$  with each other we want to draw
conclusions from non-equal probabilities only. Likewise, if $p_i$ and
$p_j$ ($1\leq i<j\leq n$) change as $p_i\to p_i+\delta$ and $p_j\to
p_j-\delta$ with a suitable (but not necessarily small) $\delta$, we
want the change of $S[p_1,...,p_n]$ to depend only on $p_i$, $p_j$
and $\delta$, but not on other probabilities $p_l$ with $l\not= i$ and
$l\not= j$; see \cite{shore,presse} for further discussion of (\ref{adon}).

\item The function $\psi(x)$ is concave: $\frac{\d^2\psi}{\d x^2}\leq
0$.  This makes $S[p_1,...,p_n]=\sum_{k=1}^n\psi(p_k)$ consistent with
(\ref{major}, \ref{schur}), because for any concave function $\psi(x)$
relations (\ref{major}) and (\ref{schur}) imply \cite{major}
\be
\no{conv}
\sum_{k=1}^n\psi(p_k) \leq \sum_{k=1}^n \psi(q_k).
\ee

\item \label{C} Instead of actions $\A_1,...,\A_n$ that refer (resp.) to
probabilities $p_1,...,p_n$, consider a composite set of actions
$\{\A_k\B_l\}_{k=1\,l=1}^{n\quad n'}$ with probabilities (resp.) $\{p_k
p'_l\}_{k=1\,l=1}^{n\quad n'}$. Hence $\{\A_k\}_{k=1}^{n}$ is independent
from $\{\B_l\}_{l=1}^{n'}$; e.g. $\{\A_k\}_{k=1}^{n}$ and
$\{\B_l\}_{l=1}^{n'}$ may refer to the same set of actions performed
independently at different times. It is natural to demand from a risk
measure $S[...]$ that it adds up for independent actions
\cite{shore,presse}:
\be
\label{ado}
S[p_1p'_1,...,p_np'_{n'}]=S[p_1,...,p_n]+S[p'_1,...,p'_{n'}].
\ee

\end{enumerate}

Conditions (\ref{A}--\ref{C}) are natural for a measure of
uncertainty|hence for a risk measure|and they lead to \cite{leod}: 
$\psi(p)=-p\ln p$, up to a positive multiplicative
constant that we fixed to $1$, and an additive constant that we fixed to
$0$ \cite{leod}. Thus the sought-after quantity amounts to the entropy
\be
\no{entrop}
\no{eq:6}
S[p]=-\sum_{k=1}^np_k\ln p_k.
\ee
Eq.~(\ref{conv}) means that entropy is larger for more risky probability. 
The expression (\ref{entrop}) for the entropy can be recovered via other
axiomatic schemes; see \cite{korner,csis} for a review.

Thus, instead of conditions (\ref{major}) we shall employ the entropy
(\ref{entrop}) as a measure of risk.\footnote{The use of entropy as a
measure of risk was criticized in literature (see e.g.,~\cite{aumann}) on
the grounds that it depends only on probability and not (also) on the
costs $\varepsilon_k$. However, we note that the criticisms of
\cite{aumann} are not general, and it certainly does not apply to the
situation we consider, where both probabilities $\{p_k\}_{k=1}^n$ and
$\{q_k\}_{k=1}^n $ refer to the same costs $\{\varepsilon_k\}_{k=1}^n$.
Indeed, the argument of~\cite{aumann} against using the entropy as a
measure of risk refers (say) to comparing two situations, where within
the first situation the agent gets $10\$$ and $-10\$$ with probabilities
$p_1$ and $p_2$. Within the second situation the agent gets $1000\$$ and
$-1000\$$ with the same probabilities $p_1$ and $p_2$.  The entropies
here will be the same, $-p_1\ln p_1-p_2\ln p_2$, but the risks are
obviously different.} Now the risk-averse agent chooses
to minimizes entropy (\ref{entrop}) under constraints
(\ref{eq:4}, \ref{eq:5}):
\be
\no{balasan} 
{\rm min}\left[
-\sum_{k=1}^np_k\ln p_k ;\quad p_1\geq ...\geq  p_n\geq 0,\quad \sum_{k}p_k\ve_k=E\right],
\ee
while the risk-seeking agent chooses in (\ref{balasan}) max instead of min. 

The maximization or minimization can be done via the Lagrange function 
\be \no{lagri} {\cal L}=
-\sum_{k=1}^np_k\ln p_k+\hat\beta \sum_{k=1}^np_k\varepsilon_k, 
\ee
where $\hat\beta$ is the Lagrange multiplier that corresponds with $E$
in (\ref{eq:4}). It is important to stress that (\ref{lagri}) emerged as
a measure of risk and uncertainty in several alternative axiomatic
schemes \cite{aczel,chin1,chin2,ng}; see Appendix \ref{ko} for a
discussion. 

\subsection{A general discussion on entropy maximization versus its minimization}
\label{maxomino}

The entropy maximization is a well-known method for determination of
prior probabilities~\cite{jaynes,shore,ttl,uffink,grun,landes1,landes2}.
This method was developed within statistical physics \cite{balian}, and
it reflects the second law of thermodynamics, i.e., the natural tendency
of the entropy to increase in closed systems \cite{balian}. The method
was also motivated from within the probabilistic inference
theory~\cite{shore,ttl,uffink}, and applied to Bayesian decision making
\cite{grun,landes1}, social group decision making~\cite{durlauf}, game
theory~\cite{wolpert}, and learning algorithms \cite{kianercy}, where it
describes bounded rational agents, etc. The entropy maximization
was also discussed from the viewpoint of approximate (and causal)
reasoning~\cite{jar_1,jar_2,jar_3,jar_4}. The maximum entropy method is
fundamental for statistics, e.g., it was recently motivated from within
the objective Bayesianism program \cite{landes1,landes2}. In social
sciences the outcome of the entropy maximization method is known as the
logit distribution; see \cite{wolpert,kianercy} for reviews. Several
basic features of entropy were related to the utility maximization in
economics \cite{rozonoer,saslow,candeal}. Mathematical psychology and
choice theory also provide interesting situations, where the agent
behavior is determined by entropy maximization for a fixed average
utility; see \cite{marley} for a thorough review. 

In contrast, the entropy minimization for a risk-averse agent is
compared with a general trend of social and biological systems that
create order, i.e.,  decrease the entropy locally~\cite{lindsay,polgar}.
Formalizations of such processes within statistical physics are less
known, but do exist~\cite{jaynes_pra,huhu,karen}. Below we shall employ
results from~\cite{karen}. Occasionally, aspects of entropy minimization
are also discussed in probabilistic inference
\cite{good,christensen,watanabe}, e.g., for the feature extraction
problem \cite{christensen}.

\comment{ One can look at gambling experiments to see whether human
subjects more frequently demonstrate risk-aversion or risk-seeking;
see~\cite{baron} for a review. Experiments show that whenever
$\varepsilon_k< 0$ (i.e.  monetary gains), people predominantly show
risk-aversion, while for $\varepsilon_k> 0$ (monetary costs) people tend
to be risk-seeking~\cite{baron}. The situation, where both negative and
positive $\varepsilon_k$'s are possible is unclear. We do not
necessarily impose definite-sign $\varepsilon_k$'s, hence we shall study
below both risk-seeking and risk-averse behavior. 

But we note that the very formulation of our problem|where the agent
should not follow the least-cost action|will be more likely percieved by
human subjects as that of costs. Formally, this is expressed by choosing
$\varepsilon_1= {\rm min}_k[\varepsilon_k]$ as a reference point and
redefining the costs as $\varepsilon_k\to \varepsilon_k-\varepsilon_1$,
so that all of them become non-negative.  Hence, it is likely that for
the considered situation many human subjects will be prone to
risk-seeking. }

%%Tomamichel

\section{Prior probabilities via entropy minimization}
\label{min}

\subsection{Parametrization of probabilities}

Here we discuss the minimization of entropy
(\ref{entrop}) under constraints (\ref{eq:4}, \ref{eq:5}); see also
\cite{karen} in this context. First, we note that (\ref{eq:5})
and (\ref{eq:4}) are compatible only for
\be
\no{manushak}
\frac{1}{n}\sum_{k=1}^n\ve_k\geq E=\sum_{k=1}^np_k\ve_k.
\ee
Indeed, we apply the summation-by-parts formula
\be
\sum_{k=1}^np_k\ve_k=\ve_n\sum_{k=1}^np_k-\sum_{m=1}^{n-1}[\ve_{m+1}-\ve_{m}]\sum_{k=1}^mp_k,
\no{abel}
\ee
to both sides of (\ref{manushak}) and obtain:
\be
\no{chuch}
\sum_{k=1}^n \frac{1}{n} \ve_k -\sum_{k=1}^np_k\ve_k=
\sum_{m=1}^{n-1}[\ve_{m+1}-\ve_{m}]\left[\sum_{k=1}^mp_k - \sum_{k=1}^m \frac{1}{n} \right],
\ee
Now (\ref{manushak}) follows, because the first (second) term inside square
brackets in (\ref{chuch}) is non-negative due to (\ref{eq:08}) (due to (\ref{borik})). 
Using (\ref{eq:08}, \ref{borik}) we parametrize the sought-after probabilities $(p_1,.., p_n)$ as~\cite{karen}
\begin{eqnarray}
  \label{eq:8}
  &&  (p_1,...,p_n)=\sum_{\alpha=1}^n \lambda_\alpha\pi_\alpha, 
  \qquad  \pi_\alpha \equiv \frac{1}{\alpha}(\underbrace{1....1}_{\alpha},
  0,...,0), \\
  &&   \label{eq:80}  \label{eq:9}
  \lambda_\alpha\geq 0, \quad \sum_{\alpha=1}^n \lambda_\alpha=1,\\  
\label{tarzan}
&& p_k=\sum_{\alpha=k}^n\frac{\lambda_\alpha}{\alpha}, \qquad k=1,...,n.
\end{eqnarray}
It is easy to show that (\ref{eq:8}) is necessary and sufficient for
holding (\ref{eq:5}); see also (\ref{eq:08}). The advantage of using
(\ref{eq:8}) is that the probabilities $\lambda_\alpha$ do not have
any other constraints besides (\ref{eq:80}) and (\ref{eq:4}) that in terms of
$\lambda_\alpha$ are written as
\begin{eqnarray}
  \label{ush}
&& E=\sum_{\alpha=1}^n\lambda_\alpha\xi_\alpha,\\
&& \xi_\alpha\equiv\frac{1}{\alpha}\sum_{i=1}^\alpha\varepsilon_i.
\label{dat}
\end{eqnarray}

\subsection{Minimization of a concave function on a convex set}

Note that constraints (\ref{borik}) and (\ref{eq:4}) define a convex set
\cite{rock}, which we denote by $\Omega$. The same convex set, but
in different coordinates, is defined via (\ref{eq:9}) and (\ref{ush}).
We now recall that the entropy $S[p]$ in (\ref{eq:6}) is a concave
function of $p=(p_1,...,p_n)$ on $\Omega$~\cite{jaynes,shore,ttl,uffink}
[cf.~(\ref{conv})]: 
\be S[\chi\, p+(1-\chi)\,q]\geq \chi\, S[p]+(1-\chi)\,S[q],
\quad 0\leq \chi\leq 1,  \no{conca} \ee 
where $q=(q_1,..., q_n)$.
Eq.~(\ref{conca}) is well-known, but it can be verified by looking at an 
unconstrained Hessian matrix (i.e., without accounting for constraints
(\ref{borik}), (\ref{eq:4})): 
\be
\no{berd}
\frac{\partial^2 S}{\partial p_k\,\partial p_l}=-\frac{\delta_{kl}}{p_k}, 
\ee
where $\delta_{kl}$ is a unit matrix. 
Once $S$ is concave without the constraints, it is also concave on $\Omega$.

Now representation (\ref{eq:8}) implies that the entropy $S$ is also a
concave function of $\lambda_\alpha$, e.g., takes derivatives over
$\lambda_\alpha$ and $\lambda_\beta$ as in (\ref{berd}). The
non-constant concave function $S[\lambda]$ (entropy in variables
$\lambda_\alpha$) on the convex set $\Omega$ can reach its local minima
only on vertices of $\Omega$---i.e.,  those elements of $\Omega$ that
cannot be represented as a convex sum of other elements of $\Omega$\footnote{\label{trivium}This fact follows from the negativity of the
Hessian matrix, but can be shown as well directly: let $\lambda_0$ be a local
minimum of $S[\lambda]$ that is not a vertex of ${\cal S}$. Then there
exist $\lambda_1$ and $\lambda_2$, both from ${\cal S}$, and both 
close to $\lambda_0$ so that $\lambda_0= \chi \lambda_1+(1-\chi)
\lambda_2$, and $0\leq \chi\leq 1$. Now $S[\lambda_0]\geq \chi
S[\lambda_1] +(1-\chi) S[\lambda_2] $ from concavity of $S[\lambda]$,
and $S[\lambda_1]>S[\lambda_0]$, $S[\lambda_2]>S[\lambda_0]$, because
$\lambda_0$ is a local minimum.  These lead to
$S[\lambda_0]>S[\lambda_0]$, which is contradictory.  Likewise, one can
show that for a concave function $S[\lambda]$ on a convex domain ${\cal
S}$, any local maximum coincides with the global one. Let $\lambda_l$
and $\lambda_g$ be, respectively, the local and global maxima of
$S[\lambda]$. Consider $S[\chi \lambda_g+(1-\chi) \lambda_l]$, where
$0<\chi< 1$ is sufficiently close to $0$.  Then $\chi \lambda_g+(1-\chi)
\lambda_l$ is close to the local maximum $\lambda_l$, but $S[\chi
\lambda_g+(1-\chi) \lambda_l]\geq S[\lambda_l]$, again is a
contradiction.}~\cite{rock}. For those vertices $\lambda_\alpha=0$ for
as many as possible indices $\alpha$.  Now for each vertex of $\Omega$,
generically only two $\lambda$'s are non-zero; otherwise (\ref{ush})
cannot hold for a given $E$. (For particular values of $E$ there can be
one non-zero $\lambda_\alpha$).  Let us take all $n(n-1)/2$ vectors
(\ref{eq:9}) with only two non-zero $\lambda_\alpha$'s that are
determined from the normalization $\sum_{\alpha=1}^n\lambda_\alpha=1$ in
(\ref{eq:9}) and from (\ref{ush}). Now when minimizing the entropy on those
vectors one finds the global entropy minimum~\cite{karen}. 

Note that the above concavity argument does not imply that all solutions
are local minima with respect to an infinitesimal perturbation that
holds (\ref{eq:4}, \ref{eq:5}).\footnote{As an example consider a sphere
intersected by a horizontal plane. The part of the plane that is inside
of the sphere defines a convex set, and the surface of the sphere is a
concave function. Vertices of the set are intersection points of the
sphere with the plane, but they are not local minima. } Nevertheless,
it is the case that for the present problem all above $n(n-1)/2$
solutions are local minima; see Appendix \ref{samson} for details. The
local minimality is important because it makes each solution
meaningful even if it does not provide the global minimum of entropy.
Note that choosing the global minimum demands ${\cal O}(n(n-1)/2)$
operations. This number scales polynomially with $n$. 

Denote by $\{\alpha\beta\}$ ($\alpha<\beta$) the solution, where only
$\lambda_\alpha\equiv\lambda_\alpha$ and $\lambda_\beta=1-
\lambda_\alpha$ are non-zero. For we get
(\ref{ush}, \ref{dat}):
\begin{eqnarray}
  \label{eq:1000}
&&    E=\lambda_\alpha\xi_\alpha
  +(1-\lambda_\alpha) 
  \xi_\beta , \qquad
  \lambda_\alpha=\frac{\xi_\beta-E}{\xi_\beta-\xi_\alpha}, \\
&& p_1=...=p_\alpha=\frac{\lambda_\alpha}{\alpha} + \frac{\lambda_\beta}{\beta}, \quad 
p_{\alpha+1}=...=p_\beta= \frac{\lambda_\beta}{\beta}, \quad 
p_{\beta+1}=...=p_n=0.
\label{menak}
\end{eqnarray}

we get the minimized entropy from (\ref{eq:6}, \ref{eq:8}, \ref{eq:9}, \ref{eq:1000}, \ref{menak})
\begin{eqnarray}
\label{eq:102}
&& S_{\{\alpha\beta\}}(E)=-\nu_{\alpha\,\beta} 
\ln\left[\frac{\nu_{\alpha\,\beta}}{\alpha}\right]-(1-\nu_{\alpha\,\beta}) 
\ln\left[\frac{1-\nu_{\alpha\,\beta }}{\beta-\alpha}\right], \\
\label{eq:103}
&& \nu_{\alpha\,\beta}\equiv\lambda_\alpha
( 1-\frac{\alpha}{\beta} )+\frac{\alpha}{\beta} =\frac{\xi_\beta-E}{\xi_\beta-\xi_\alpha}
\,( 1-\frac{\alpha}{\beta} )+\frac{\alpha}{\beta} .
\end{eqnarray}
Now note from (\ref{eq:08}, \ref{dat}) that [for $\alpha<\beta$]
\begin{eqnarray}
  \xi_\alpha-\xi_\beta&=&\frac{\beta-\alpha}{\alpha\beta}
\sum_{i=1}^\alpha\varepsilon_i
-\frac{1}{\beta}\sum_{i=1+\alpha}^\beta\varepsilon_i\leq 
\frac{\beta-\alpha}{\alpha\beta}
\sum_{i=1}^\alpha\varepsilon_i - \frac{\beta-\alpha}{\beta}
\,\varepsilon_{\alpha+1} =
\frac{\beta-\alpha}{\alpha\beta}
\sum_{i=1}^\alpha [\,\varepsilon_i - \varepsilon_{\alpha+1}
\,]\leq 0. 
  \label{eq:11}
\end{eqnarray}
Hence, within solution $\{\alpha\beta\}$ ($\alpha<\beta$)
we get [see
(\ref{eq:1000})]
\begin{eqnarray}
  \label{eq:13}
\xi_\alpha\leq E\leq \xi_\beta. 
\end{eqnarray}
Recall from (\ref{manushak}) that the allowed range of $E$ is 
$\varepsilon_1=\xi_1\leq E\leq \xi_n$.

\subsection{Features of the minimized entropy}

We obtain from (\ref{dat}--\ref{eq:11})
\begin{eqnarray}
  \frac{\d S_{\{\alpha\beta\}} }{\d E}&=&\frac{\partial\nu_{\alpha\beta} }{\partial E}\,\,
\frac{\partial S_{\{\alpha\beta\}} }{\partial\nu_{\alpha\beta} }
=\frac{1-\frac{\alpha}{\beta}}
  {\xi_\beta-\xi_\alpha}~\ln\left[
    \frac{\nu_{\alpha\,\beta}}{1-\nu_{\alpha\,\beta}}
    \,\, \frac{\beta-\alpha}{\alpha}
  \right]=\frac{1-\frac{\alpha}{\beta}}
  {\xi_\beta-\xi_\alpha}~\ln\left[1+\frac{\beta}{\alpha}~
    \frac{\xi_\beta-E}{E-\xi_\alpha}\right]
  \geq 0,
  \label{eq:12}
\end{eqnarray}
where (\ref{eq:12}) follows from (\ref{eq:13}). According to
(\ref{eq:12}), $S_{\{\alpha\beta\}}$ is an increasing function of $E$
within each solution ${\{\alpha\beta\}}$, hence also for the
global minimum. Likewise, one can show from (\ref{eq:12}) that
\be
\no{star}
\frac{\d^2 S_{\{\alpha\beta\}} }{\d E^2}<0, 
\ee
i.e., $S_{\{\alpha\beta\}}(E)$ is concave within each solution. Hence the global minimum
${\rm min}_{1\leq \alpha<\beta\leq n}[\, S_{\{\alpha\beta\}} (E)\,]$ is
also a concave function of $E$. 

Eq.~(\ref{eq:13}) shows that different local solutions exist for
different values of $E$. Now solutions $\{\alpha\beta\}$
($\alpha>\beta$) and $\{\beta\gamma\}$ ($\beta<\gamma$) can be glued
together at $E=\xi_\beta$ so that the probabilities (\ref{menak}) behave
continuously at $E=\xi_\beta$; cf.~(\ref{eq:1000}). Indeed, at
$E=\xi_\beta$ we get $p_1=...=p_\beta=\frac{1}{\beta}$,
$p_{\beta+1}=...=p_n=0$, both for $\{\alpha\beta\}$ and
$\{\beta\gamma\}$. The gluing operation is denoted by $\vee$. The
purpose of gluing is that local solutions are combined to cover the
whole possible range $\varepsilon_1=\xi_1\leq E\leq \xi_n$ (of $E$) 
via continuous probabilities. For
$n=3$ there are two global solutions: $\{12\}\vee \{23\}$ and $\{13\}$.
The entropies are given as
\begin{equation}
 S_{\{12\} \vee \{ 23\}}(E)=
\begin{cases}
S_{\{12\}}(E)\quad  {\rm for} \quad  \xi_1\leq E\leq \xi_2, \\
S_{\{23\}}(E)\quad  {\rm for}\quad   \xi_2\leq E\leq \xi_3, 
  \end{cases}
\label{eq:14}
\end{equation}
and where $S_{\{\alpha\beta\}}$ is given by (\ref{eq:102}).  For $n=4$ we
have: $\{12\} \vee \{ 23\}\vee \{ 34\}$, $\{12\}
\vee \{ 24\}$, $\{13\}\vee \{ 34\}$, and $\{14\}$.  

Within solution $\{1n\}$ (e.g., $\{13\}$ for $n=3$) the action with the
lowest cost $\varepsilon_1$ is assigned the largest probability, while
{\it all} other actions have the same probability; see (\ref{eq:8}).
This is indeed the simplest possible prescription that provides non-zero probabilities for all utilities and holds for all possible
values of $E$. The local minimum solution $\{1n\}$ coincides with the
$\epsilon$-greedy scheme from reinforcement learning~\cite{barto}.
It is seen to correspond to (at least) a local minimum of entropy. The
candidate solution $\{12\}$ refers to the situation, where only the best
and the second-best actions are assigned non-zero probabilities. The
validity range for this local minimum is restricted by
$\xi_1=\varepsilon_1\leq E\leq \xi_2=\frac{\varepsilon_1+
\varepsilon_2}{2}$. 

Thus within the present set-up these two simple schemes of assigning
non-deterministic prior probability appear naturally. But in contrast to
imposing them {\it ad hoc}, these schemes now have their validity
conditions. 

%They are weakly (strongly) valid once they are local (global) minima
%of entropy. 

\section{Entropy maximization for risk-seeking agents}
\no{maxent}

Within the maximum entropy scheme, the risk-seeking agent will maximize
over probabilities $p_k$ the entropy (\ref{eq:6}) under constraint
(\ref{ush})~\cite{jaynes,shore,ttl,uffink}. This maximization starts from (\ref{lagri}), and it
produces the Gibbs-Boltzmann probabilities
\cite{jaynes,shore,ttl,uffink,grun,landes1,landes2} 
[cf.~Footnote \ref{trivium}]:
\begin{eqnarray}
  \label{si}
\hat p_k=\frac{1}{Z}\,e^{-\hat\beta \varepsilon_k},\qquad Z\equiv\sum_{l=1}^n
  e^{-\hat\beta \varepsilon_l},
\end{eqnarray}
where $\hat\beta$ is the Lagrange factor determined from $E=\sum_{k=1}^n\hat
p_k\varepsilon_k$; cf. (\ref{eq:4}, \ref{lagri}). The sign of $\hat\beta$ coincides
with the sign of $\frac{1}{n}\sum_{k=1}^n\ve_k-E$; see (\ref{manushak}).
Hence $\hat\beta\geq 0$, since we assume the validity of
(\ref{manushak}).  In statistical mechanics\footnote{Probabilities in
statistical mechanics can have an objective meaning; e.g., the
Gibbs-Boltzmann probabilities (\ref{si}) can refer to energy occupations of a
system in contact with a much larger thermal bath at temperature $1/\hat\beta$
\cite{balian}. But they can also have a subjective meaning---e.g., when
making predictions about a complex, closed system \cite{balian}. }
$\hat\beta$ refers to inverse (absolute) temperature, and its positivity is a
well-known fact \cite{balian}. 

It is not widely known in non-physical communities that the
Gibbs-Boltzmann probabilities (\ref{si}) have specific responses to
parameters, which were uncovered before statistical mechanics
\cite{balian}. Below we recall some features of (\ref{si}) and compare
them with the entropy minimization scheme.\footnote{An alternative way
of obtaining (\ref{si}) is to fix the entropy (\ref{eq:6}) to a non-zero
value $\hat S$, and then minimize the average cost $\sum_{k=1}^n p_k\,
\varepsilon_k$. Now $\hat\beta$ in (\ref{si}) is defined from a fixed
value of entropy. In the context of our problem, this way of looking at
(\ref{si}) is not adequate, since for us it is important that $E$ is an
independent variable that can be decided by the agent. However, in a
different setting this way of looking at (\ref{si}) led to a definition
of risk~\cite{chisar}.} 

Note that the probabilities at the local entropy minimum $\{\alpha\beta\}$
are not susceptible to changes of $\varepsilon_\gamma$ (with
$\gamma>\beta$) that take place under a constant $E$; see
(\ref{eq:1000}--\ref{eq:103}). This differs from the Gibbs-Boltzmann
probabilities (\ref{eq:6}) for the risk-seeking agent, where changing
any cost $\varepsilon_k$ under a fixed $E$ will change $\hat\beta$:
\begin{eqnarray}
  \label{eq:17}
  \left.\frac{\partial \hat\beta}{\partial \varepsilon_k}\right|_E= 
\frac{\hat{p}_k[1+\hat\beta(E-\varepsilon_k)]}{
  \sum_{l=1}^n\hat{p}_l(\varepsilon_l-E)^2},
\end{eqnarray}
where the derivative is taken under a constant $E$. Hence all the
probabilities $\{\hat p_l\}_{l=1}^n$ will generally change including
when changing $\varepsilon_k$. Ratios $\hat p_l/\hat p_m
=e^{-\hat\beta (\varepsilon_l-\varepsilon_m)}$, where $l\not= k$ and
$m\not= k$, will change as well. This corresponds with the fact that there is a
single maximum entropy solution (\ref{si}), while the entropy
minimization produces several competing local minima. 

Another difference is that in the maximum entropy
situation, as seen from (\ref{si})
\be
\no{bru}
\ve_k\approx\ve_l ~~{\rm implies}~~\hat p_k\approx\hat p_l.
\ee
For the entropy minimizing situation, (\ref{bru}) is not neccesarily the case,
as implied by (\ref{menak}), and as seen more explicitly below in 
(\ref{katon1}--\ref{katon3}). Due to this feature the entropy minimizing
solution is capable of detecting even small differences between the costs.
This is also one reason we insisted on strict inequalities in (\ref{eq:08}).

The maximized entropy is expressed from (\ref{si}) as
\be
\no{tu} 
\hat S=-\sum_{k=1}^n \hat p_k\ln \hat p_k=\hat\beta E+\ln Z.
\ee
Differentiating (\ref{tu}) and employing 
\be\no{buratino}
\frac{\d\hat\beta}{\d E}=-\frac{1}{{\sum_{k=1}^n \hat p_k(\ve_k-E)^2  }},
\ee
obtained from $\sum_{k=1}^n \hat p_k\ve_k=E$, we get
\be
\no{hasan}
\frac{\d \hat S}{\d E}=\hat\beta.
\ee
Eq.~(\ref{hasan}) shows that for $\hat\beta\geq 0$
(which holds due to (\ref{manushak})), $\hat S$ is an increasing function of $E$; cf.~(\ref{eq:12}).
Likewise, one shows that $\hat S(E)$ is a concave function of $E$:
$\frac{\d^2 \hat S}{\d E^2}\leq 0$; cf. (\ref{star}). Now using (\ref{tu}, \ref{eq:17}) we find
\begin{eqnarray}
  \label{eq:177}
  \left.\frac{\partial \hat S}{\partial \varepsilon_i}\right|_E =
-\hat\beta \hat p_i.
\end{eqnarray}
It is seen that
(\ref{eq:177}) is negative for $\beta\geq 0$; cf.~(\ref{manushak}).
Hence the uncertainty $\hat{S}$ decreases if one of the costs increases
for a fixed $E$. A similar feature holds as 
well for the minimized entropy in (\ref{eq:102}): 
\be
\left.\frac{\partial S_{\{\alpha\beta\}} }{\partial \ve_\gamma}\right|_E=
\left.\frac{\partial\nu_{\alpha\beta} }{\partial \ve_\gamma}\right|_E
\,\,
\frac{\partial S_{\{\alpha\beta\}} }{\partial\nu_{\alpha\beta} }=
(1-\frac{\alpha}{\beta})\,\,\left.\frac{\partial\lambda_{\alpha} }{\partial \ve_\gamma}\right|_E\,\,
\frac{\partial S_{\{\alpha\beta\}} }{\partial\nu_{\alpha\beta} }\leq 0.
\label{comrad}
\ee
Inequality (\ref{comrad}) follows from $\alpha<\beta$ (by definition of
$S_{\{\alpha\beta\}}$), from $\left.\frac{\partial\lambda_{\alpha}
}{\partial \ve_\gamma}\right|_E\geq 0$ (which can be shown easily from
(\ref{eq:1000})), and from $\frac{\partial S_{\{\alpha\beta\}}
}{\partial\nu_{\alpha\beta} }\leq 0$, which is seen from (\ref{eq:12}).
Eqs.~(\ref{eq:177}, \ref{comrad}) imply an interesting feature of
entropy optimization that we explore below in more detail. A large cost
$\varepsilon_\gamma$ tends to be relevant (irrelevant) for entropy
minimization (maximization), since it decreases the entropy. 

Using (\ref{si}, \ref{buratino}) one can show that 
\be\no{nono}
  \left.\frac{\partial \hat p_k}{\partial E}\right|_{\{\ve_i\}_{i=1}^n} =
-\frac{\d\hat\beta}{\d E}\,\hat p_k\,(\ve_k-E).
\ee
Due to $\frac{\d\hat\beta}{\d E}\leq 0$, we get from (\ref{nono}) that
the probabilities of sufficiently high ($\ve_k>E$) costs increase upon
increasing $E$. This feature is both natural and expected: the more utility
one is ready to invest (into adaptation), the more higher-costs actions he
explores. We shall see below that it need not hold for the
entropy-minimizing agent.

Finally, we recall that there is a
well-known freedom associated with the definition of costs
$\varepsilon_k$~\cite{luce}:
\be
\label{freedom}
\varepsilon_k\to a\varepsilon_k+b, 
\ee
where $a>0$ and $b$ are arbitrary. In particular, (\ref{freedom}) means that
probabilities assigned to a given set of actions should not change, if
all utilities (of those actions) are multiplied by a common positive
factor $a$, and/or are shifted by a common arbitrary number $b$. It is
clear that the probabilities for both entropy minimization and entropy
maximization do hold (\ref{freedom}); see in this context
(\ref{eq:1000}, \ref{eq:102}, \ref{eq:103}) and (\ref{si}) and note that
under (\ref{freedom}), $E$ also changes as $E\to aE+b$.

\section{Examples}
\no{simplest}\no{compa}

\subsection{Probabilities for three actions ($n=3$)}

Let us now study in detail the $n=3$ situation (\ref{eq:14}): this is
the simplest case that illustrates the difference between maximizing
entropy and minimizing it. Indeed, for $n=2$ the probabilities $p_1$ and
$p_2=1-p_1$ are uniquely determined by the contraint (\ref{eq:4}).
Eqs.~(\ref{eq:8}, \ref{eq:1000}) imply
\begin{align}
  \label{eq:18}
& \{12\}: (p_1,p_2,p_3)=\left(\frac{1}{2}+\frac{\lambda_{\{12\}}}{2},\,
\frac{1}{2}-\frac{\lambda_{\{12\}}}{2},\,0   \right), & 
\lambda_{\{12\}}=\frac{\xi_2-E}{\xi_2-\xi_1}, \\
  \label{eq:181}
& \{23\}: (p_1,p_2,p_3)=\left(\frac{1}{3}+\frac{\lambda_{\{23\}}}{6},\,
    \frac{1}{3}+\frac{\lambda_{\{23\}}}{6},\, 
    \frac{1}{3}-\frac{\lambda_{\{23\}}}{3}   
  \right), & 
  \lambda_{\{23\}}=\frac{\xi_3-E}{\xi_3-\xi_2}, \\
& \{13\}: (p_1,p_2,p_3)=\left(\frac{1}{3}+\frac{2\lambda_{\{13\}}}{3},\,
    \frac{1}{3}-\frac{\lambda_{\{13\}}}{3},\, \frac{1}{3}
   -\frac{\lambda_{\{13\}}}{3}   
  \right), &
  \lambda_{\{13\}}=\frac{\xi_3-E}{\xi_3-\xi_1}.
  \label{eq:182}
\end{align}
Recall that within $\{12\}$ only the two smallest costs get non-zero
probabilities; $\{23\}$ prescribes equal probabilities to those two
smallest costs, while $\{13\}$ gives equal probabilities to the two
highest cost actions.

Using the freedom provided by (\ref{freedom}), we fix $\varepsilon_1=0$
and $\varepsilon_3=1$; hence $0<\varepsilon_2<1$. Now (\ref{eq:18}--\ref{eq:182})
simplify as follows:
\begin{align}
  \label{katon1}
& \{12\}: (p_1,p_2,p_3)=\frac{1}{\ve_2}\left(\ve_2-E,E,0  \right), && 0\leq E\leq \frac{\ve_2}{2}, \\
%& \qquad \quad S_{\{12\}}=-\left(1-\frac{E}{\ve_2} \right)\ln \left(1-\frac{E}{\ve_2} \right) -\frac{E}{\ve_2}\ln \frac{E}{\ve_2},\\
  \label{katon2}
& \{23\}: (p_1,p_2,p_3)=\frac{1}{2-\ve_2}\left(1-E,1-E,2E-\ve_2\right),   && \frac{\ve_2}{2}\leq E\leq \frac{1+\ve_2}{3}, \\
%&  \qquad \quad S_{\{23\}}= -\left(1-\frac{2E-\ve_2}{2-\ve_2} \right)\ln\left[\frac{1}{2} \left(1-\frac{2E-\ve_2}{2-\ve_2} \right)\right]
%-\frac{2E-\ve_2}{2-\ve_2} \ln\frac{2E-\ve_2}{2-\ve_2}, \\
& \{13\}: (p_1,p_2,p_3)= \frac{1}{1+\ve_2}\left(1+\ve_2-2E,E,E  \right), && 0\leq E\leq \frac{1+\ve_2}{3}.
%&\qquad \quad S_{\{13\}}= -\left(1-\frac{2E}{1+\ve_2} \right)\ln \left(1-\frac{2E}{1+\ve_2} \right) -\frac{2E}{1+\ve_2}\ln \frac{E}{1+\ve_2}.
  \label{katon3}
\end{align}
It is seen that (\ref{katon2}) and (\ref{katon3}) do not coincide with 
each other for $\ve_2\to \ve_1+0$ ($\ve_1=0$);
cf. the discussion around (\ref{bru}).

Note that within each solution (\ref{katon1}--\ref{katon3}) increasing
$E$ leads to larger probabilities for costly actions 2 and 3. We shall
show below that this intuitively expected feature does not hold for
transitions between different solutions (i.e., local minima of entropy). 

For various parameter regimes we shall now compare the behavior of the
risk-averse (entropy-minimizing) agent with the risk-seeking
(entropy-maximizing) one. Everywhere, such comparisons are made for the
same values of $\{\varepsilon_k\}_{k=1}^n$ and the same value of $E$
(where $E$ is the utility invested into exploration). As seen below,
entropy-minimizing probabilities are neither majorized nor majorized by the
entropy-maximizing ones, i.e., (\ref{tatar}) and (\ref{abver}) are valid.
This fact makes the situation interesting, in particular because (for
$n=3$) there are two strategies for risk-aversion: putting a larger
weight on the lowest-cost action or a lesser weight on the highest-cost
action; cf. our discussion after (\ref{schur}).

\subsection{Two low-cost actions}

\begin{figure}
  \centerline{\includegraphics[width=.4\textwidth]{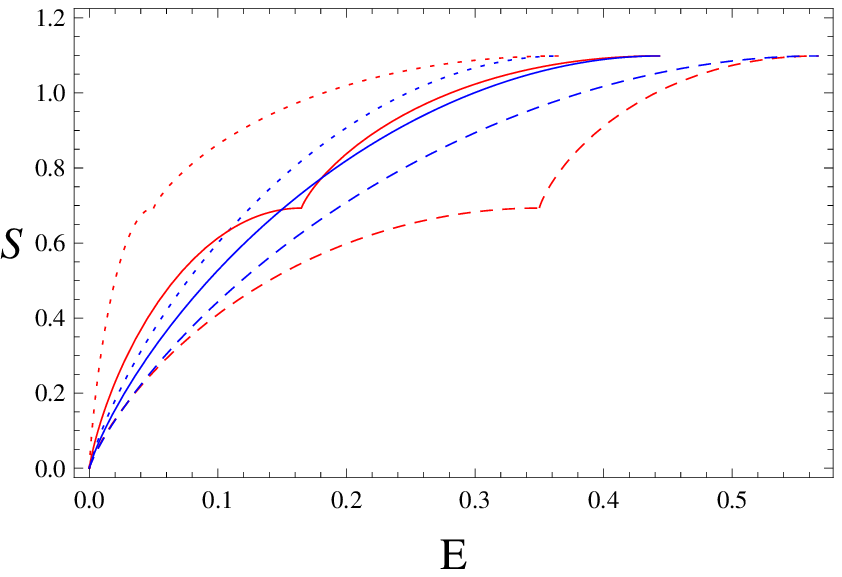}}
  \caption{(Color online) The locally minimal entropies $S_{\{12\} \vee
    \{ 23\}}(E)$ (red curves) and $S_{\{13\}}(E)$ (blue curves) as
functions of $E$ for $n=3$; see (\ref{eq:14}).
Using (\ref{freedom}) we fixed for costs: $\varepsilon_1=0$,
$0<\varepsilon_2<1$, and $\varepsilon_3=1$.\\
    Dotted curves: $\varepsilon_2=0.1$. The blue curve is always lower: $S_{\{12\} \vee
    \{ 23\}}(E)>S_{\{13\}}(E)$; hence the local minimum $\{13\}$ is the global entropy minimum. \\
    Full curves: $\varepsilon_2=0.33$. Now the blue and red curves intersect
    two times: the local minimum $\{13\}$ is not the global one for $0.15\leq E\leq 0.18$. 
    The transition from $\{12\}$ to $\{23\}$ takes place at $E=0.165$. \\
    Dashed curves: $\varepsilon_2=0.8$. Now the red curve is always
    lower, $S_{\{12\} \vee \{ 23\}}(E)<S_{\{13\}}(E)$, meaning that the solution $\{12\} \vee \{ 23\}$ is the global minimum. }
\label{fig1}
\end{figure}

In this regime two lower costs are approximately equal:
$\varepsilon_1\approx\varepsilon_2$.  Fig.~\ref{fig1} displays the
entropies $S_{\{12\} \vee \{23\}}(E)$ and $S_{\{13\}}(E)$ for a
representative set of parameters; see (\ref{eq:14}). It is seen that for
$\varepsilon_1\approx\varepsilon_2$ the global minimum is always
$\{13\}$; cf. Fig.~\ref{fig1}. Hence the corresponding probabilities are
given by (\ref{katon3}), where two actions with non-minimal costs
$\ve_2$ and $\ve_3$ are given the same weight. 

\begin{figure}
  \centerline{\includegraphics[width=.4\textwidth]{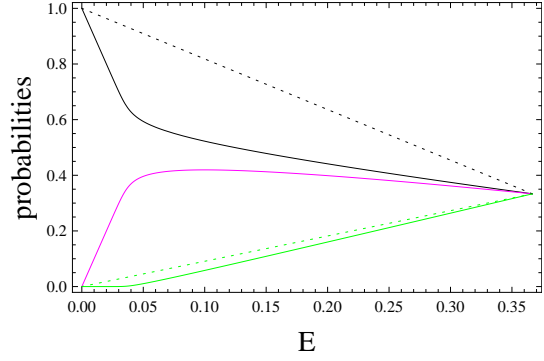}}
  \caption{(Color online) Probabilities as a function of $E$ for the
    $n=3$ situation with $\varepsilon_1=0$, $\varepsilon_2=0.1$ and
    $\varepsilon_3=1$. Full curves: $\hat{p}_1$ (black), $\hat{p}_2$
    (magenta), $\hat{p}_3$ (green). Dotted curves: $p_1$ (black),
    $p_2=p_3$ (green). \\
    The Gibbs-Boltzmann probabilities $\hat{p}_k$
    for the risk-seeking agent are calculated from (\ref{si}).
    The probabilities $\{p_k\}_{k=1}^3$ refer to $\{13\}$ [see (\ref{katon3})], which
    is the global minimum for the present values of $\varepsilon_k$;
    cf. Fig.~\ref{fig1}.  }
\label{fig2}
\end{figure}

In the regime $\{13\}$ the entropy minimizing probabilities
(\ref{katon3}) hold [cf. Fig.~\ref{fig2}]
\begin{eqnarray}
\label{a13}
\{13\}:
&& p_1>\hat{p}_1,\\
&& p_3>\hat{p}_3,
\label{b13}
\end{eqnarray}
where $0< E< \frac{1+\ve_2}{3}$, and $\hat p_k$ are the Gibbs-Boltzmann
probabilities (\ref{si}). Hence $(p_1, p_2, p_3)$ neither majorizes
$(\hat p_1, \hat p_2, \hat p_3)$ nor is majorized by that, i.e., (\ref{a13},
\ref{b13}) amount to a particular case of (\ref{abver}). 

Now (\ref{a13}) means that the entropy-minimizing agent invests more
probability on the lowest-cost action than the entropy-maximizing one.
This is one strategy for risk-aversion.\footnote{
We can look at (\ref{a13}) from a different viewpoint. Recall that for $E\to\ve_1=0$ both
entropy-minimization and entropy-maximization produce: $p_1\to 1$ and
$\hat p_1\to 1$, i.e., they converge (as they should) to taking the
least-cost action. Now one can ask to which extent this
least-cost action is stable $(\widetilde p_1, \widetilde p_2, \widetilde
p_3)=(1,0,0)$ with respect to a small, but non-zero $E$. This question
can be addressed by looking at one of standard
distances between probabilities, e.g.,  the variational distance
$\Delta_1[p(E),\widetilde
p]\equiv\frac{1}{2}\sum_{k=1}^3|p_k(E)-\widetilde p_k|$ or the Hellinger
distance $\Delta_2[p(E),\widetilde p]\equiv 1-\sum_{k=1}^3\sqrt{
p_k(E)\,\widetilde p_k}$. It is clear that $\Delta_1[\hat
p(E),\widetilde p]-\Delta_1[p(E),\widetilde p]=p_1-\hat p_1$ and
$\Delta_2[\hat p(E),\widetilde p]-\Delta_2[p(E),\widetilde
p]=\sqrt{p_1}-\sqrt{\hat p_1}$. Now (\ref{a13}) and Fig.~\ref{fig1} show that for a
small but non-zero $E$ we have $\Delta_k[\hat p(E),\widetilde
p]>\Delta_k[p(E),\widetilde p]$ ($k=1,2$), i.e. the least-cost action is
more stable for the entropy minimizing agent.} Inequality (\ref{b13}) means that the
entropy-minimizing agent assigns to costly actions
more probability than the entropy-maximizing agent. Note however that in the
considered regime $\varepsilon_1\approx\varepsilon_2$, where the
solution $\{13\}$ is the global minimum of energy we have (in addition
to (\ref{b13})) 
\be
\no{kar}
p_1-\hat p_1\gg p_3-\hat p_3, 
\ee
except $E=0$ and $E=\frac{1+\ve_2}{3}$, where $p_1-\hat p_1=p_3-\hat
p_3=0$; cf.  Fig.~\ref{fig2}. However, if $\ve_2$ is sufficiently
larger than $\ve_1=0$, relation (\ref{kar}) need not hold---e.g., for $0.51<
\ve_2<0.68$ and sufficiently small $E$, the global minimum of entropy is
still given by $\{13\}$. There we can have (together with (\ref{a13},
\ref{b13})): $p_3-\hat p_3- (p_1-\hat p_1)\simeq 0.01>0$; i.e.,
though the difference is relatively small it is still positive. 

Another difference is that the entropy-minimizing (maximizing) agent
tends to underweight (overweight) the middle-cost action:
\be
\no{jimmi}
1=\frac{p_3}{p_2}>\frac{\hat p_3}{\hat p_2}. 
\ee
The entropy-maximizing agent focuses on this middle-cost action and
ignores the most costly action since it has $\hat{p}_1\simeq
\hat{p}_2\gg \hat{p}_3$; cf. Fig.~\ref{fig2}.

Thus, the feature of $\{13\}$ is that it puts more weight on the
lowest-cost action|which is one possibility of risk-aversion|and does
account for highest-cost actions in the sense of (\ref{b13},
\ref{jimmi}).

\subsection{Two high-cost actions: distinguishing between actions with approximately equal costs}

\begin{figure}
  \centerline{\includegraphics[width=.4\textwidth]{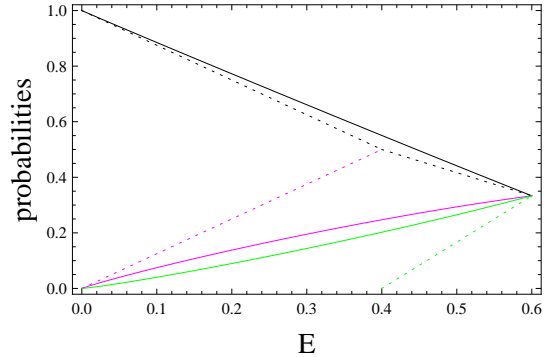}}
  \caption{(Color online) The same as in Fig.~\ref{fig2}, but now
    $\varepsilon_1=0$, $\varepsilon_2=0.8$ and
    $\varepsilon_3=1$. Dotted curves refer to the solution $\{12\}
    \vee \{ 23\}$ [see (\ref{katon2})], which is now the
    global minimum: $p_1$ (black), $p_2$ (magenta; note that $p_2=p_3$
    for $E\geq \xi_2=\frac{\ve_2}{2}=0.4$), and $p_3$ (green; we have $p_3=0$ for
    $E\leq \xi_2=\frac{\ve_2}{2}=0.4$).}
\label{fig3}
\end{figure}

Let us now study the opposite case $\varepsilon_1\ll\varepsilon_2\approx
\ve_3=1$.  For a sufficiently small $E$ the global minimum of
entropy is $\{1 2\}$; see Fig.~\ref{fig1}. For the solution $\{12\}
\vee \{23\}$ the inequalities (\ref{a13}, \ref{b13}) are inverted:
\begin{eqnarray}
\label{a1223}
\{12\} \vee \{23\}:
&& p_1<\hat{p}_1,\\
&& p_3<\hat{p}_3. 
\label{b1223}
\end{eqnarray}
In this regime another strategy of risk-aversion is realized---the
entropy-minimizing agent puts less weight on the highest-cost action. In
particular, for $0\leq E\leq \frac{\ve_2}{2}$ this action gets no
probability at all: $p_3=0$; see (\ref{eq:18}). Hence when having two
possibilities $\ve_2$ and $\ve_3$ with comparable high costs
$\ve_2\lesssim\ve_3$, the agent prefers the lesser of two evils. In
contrast, the entropy-maximizing agent will take these options with
nearly equal probabilities. For larger values of $E$, the global minimum
is $\{2 3\}$, where all probabilities are non-zero, but the
probability for the action with the (intermediate) cost $\ve_2$ is as
large as for the least-cost action: $p_2=p_1$; see (\ref{eq:181}) and
Figs.~\ref{fig1} and \ref{fig3}. 

We see here an effect whose traces were also observed in the previous
scenario: when choosing between two actions of comparable cost
$\ve_2\approx\ve_3$, the entropy-minimizing agent is able to distinguish
between them despite a small difference. In contrast to this, the
entropy-maximizing agent will just take them with (approximately) equal
probability, neglecting that small difference. 

%Thus we can say that the entropy-minimizing agent is cautious, while the
%entropy-maximizing one is risky and careless. 

\subsection{Transitions from one regime to another, re-entrance,
cognitive dissonance and frustration}

\subsubsection{Transition between $\{13\}$ and $\{12\}$ upon increasing 
$E$}

So far we studied cases $\varepsilon_1\approx\varepsilon_2\ll \ve_3=1$
and $\varepsilon_1\ll\varepsilon_2\approx \ve_3=1$, where one (global)
solution|respectively, $\{13\}$ and $\{12\}\vee \{23\}$|provides the
global entropy minimum for all values of $E$. Now we turn to studying
cases, where $\varepsilon_1\not\approx\varepsilon_2$ and
$\varepsilon_3\not\approx\varepsilon_2$. Here it is possible to have
transitions between different local minima upon changing $E$; see
Fig.~\ref{fig1} with the case $\ve_2=0.33$. 

For a sufficiently small $E$ 
the global minimum is $\{1 3\}$. But for $E>0.15$, the global minimum
becomes $\{1 2\}$. This transition from one regime to another is
continuous in terms of entropy, but discontinuous in terms of probabilities
for various actions, as seen from (\ref{eq:18}--\ref{eq:182}). This is
an analogue of the first-order phase transition in statistical physics
systems (e.g., the liquid-vapor transition) \cite{balian}, where the role
of $E$ is played by the (physical) temperature.  There the thermodynamic
potential whose minimization determines stability (for our case this is
entropy) changes continuously, but the order parameter (the difference
between densities of liquid and vapor) suffers a discontinuous change
\cite{balian}. 

The transition at $E= 0.15$ is against naive intuition, because with a
larger average cost $E$ the agent neglects the action $\ve_3$, which is
related to the highest cost. Here the agent who invests less into
adaptation assigns more probability to costly actions. For the
entropy-maximizing agent it is impossible that the probability of the
most costly action decreases upon increasing $E$; cf. our discussion
around (\ref{nono}). For the entropy-minimizing agent such an effect
does take place|the probability $p_3$ of the most costly state changes
from a positive value to zero|due to transition from one local minimum
to another. For changes within the local minimum this cannot happen, as
seen from (\ref{katon1}--\ref{katon3}). 
This effect can be related to cognitive
dissonance~\cite{disso,aronson,akela} \footnote{It is characterized by 
the following feature: the more energy and/or effort people invest
into some situation, the narrower the set of their actions or intentions
tends to become. (This narrowing was described theoretically
within probabilisitic opinion formation \cite{AG}.) E.g. the more
(possessions and/or time) people invest into a sectarian movement, the more
vigorously they tend to support it~\cite{disso,aronson,akela}. Another
example: once people already buy something, they 
tend to have fewer doubt about its value and relevance. The 
above behavior of the entropy-minimizing agent agrees with cognitive
dissonance---the more utility the agent decides to invest into adaptation
(i.e., $E$ increases), and the narrower his action set tends to become. 
The underlying cause of this effect in our situation also relates to 
one of the phenomenological features of cognitive dissonance
\cite{disso,aronson}; that is, people tend to minimize the uncertainty in their
action (belief or intention) set, which for our situation refers to
minimizing entropy (uncertainty).  Within the cognitive dissonance
theory these two aspects---namely that investing more means believing more 
and uncertainty minimization---are related to each other other on the grounds
of a general plausibility \cite{disso,aronson}.  Here we see that they
are related to each directly, and risk
minimization explains this relation. }.

\comment{Cognitive dissonance is characterized (among others) by
following features: {\it (i)} people do not support widely different
opinions, intentions or beliefs. They tend to get rid most of one of
them in favour of one (opinion, intention or belief). {\it (ii)} The
more energy and/or effort is invested into the opinion (intention or
belief) formation, the more narrow it tends to become \footnote{For
example, more (possesions and/or time) people invest into a sectarian
movement, more vigorously they tend to support it~\cite{disso}.}.
Analogues of these features are seen above: under {\it increasing} the
invested utility $E$, the agent moves abruptly from one set of
probabilities to another, e.g.  from $\{1 3\}$ to $\{1 2\}$.
Moreover, the latter regime is more narrow in the sense that the
probability of the action with cost $\ve_3$ is explicitly zero: $p_3=0$.
In this context, recall the analogy between the cost and energy; cf.
after (\ref{eq:5}). Thus we conclude that the entropy-minimizing agent
does show features of cognitive dissonance. }

\subsubsection{Re-entrance}

However, the above transition from $\{13\}$ to $\{12\}$ upon increasing $E$ is 
not the end of the story: at $E=0.165$ (and $\ve_2=0.33$) the
solution $\{12\}$ continuously (both in terms of entropy and
probabilities) changes to $\{2 3\}$. This is a natural transition: for
$E>0.165$, the solution $\{1 2\}$ does not exist anymore, since it
cannot hold the constraint $E=p_2\ve_2$; cf.~(\ref{eq:13}). This is an
analogue of a second-order phase transition in statistical physical
systems (e.g., paramagnetic-ferromagnetic transition in magnets)
\cite{balian}, where one solution ceases to exist and is replaced by
another via a continuous change of the order parameter (magnetization,
in our case probabilities of actions).\footnote{We emphasize that a
normal understanding of phase-transitions relates them to the case $n\to
\infty$. This condition is however not necessary for having a
phase-transition, e.g., in a micro-canonical ensemble \cite{balian}.} 

But eventually, for $E>0.18$, the global minimum is back from $\{2
3\}$ to $\{1 3\}$ (re-entrance).  The probabilities of the solution
$\{1 3\}$ stay almost constant in the whole interval $0.15\leq E\leq
0.18$. Note that the re-entrance effect persists up to $\ve_2\leq 0.45$.
For $0.68>\ve_2>0.45$ (not shown on figures) the re-entrance behavior is
absent; there is only a single transition from $\{1 3\}$ to $\{1
2\}$ upon increasing $E$.  Eventually, for $0.68<\ve_2$ the global
entropy minimum is always $\{1 2\}\vee \{2 3\}$, and we revert to
the studied regime $\varepsilon_1\ll\varepsilon_2\approx \ve_3=1$. 

The transitions from $\{13\}$ to $\{12\}\vee\{23\} $ and back mean that
there are three local minima with comparable values of entropy but
different values for probabilities of actions; cf.~Fig.\ref{fig1}. They
compete with each other upon relatively small changes of $E$. This
effect resembles frustration, whose psychological content is that there are 
two (or more) different (incommensurate) goals or motivations that
compete with one other. The concept of frustration is also well-known
in statistical physics of complex systems (see \cite{kats} for a recent
review), where its meaning is very close to the above, since it relates
to competing local minima of the thermodynamic potential (entropy in
our case). 

We stress that neither of the above effects is seen for the risk-seeking
agent, because in this situation probabilities (\ref{si}) are unique and
depend smoothly on the parameters involved. This has to do with the fact
that (\ref{si}) was obtained via maximization of a concave function in a
convex set, which generically produces unique and well-behaved
results~\cite{rock}. Possible non-uniqueness of the Gibbs-Boltzmann
probabilities (\ref{si}) can show up only in the thermodynamic limit
$n\to\infty$ for specific systems that can be subject to
phase-transitions. (We however refrain from considering the $n\to\infty$
case, since it is far from an agent facing a limited amount of different
choices.)

\subsection{Four actions}

We turn to the entropy-minimizing scenarios for $n=4$. We do so briefly,
because though $n=4$ is richer then $n=3$, it does not offer conceptual
novelties.  Here we also fix $\ve_1=0$ and $\ve_4=1$
[cf.~(\ref{freedom})], and we are left with two parameters: $0<\ve_2<
\ve_3<1$.  For $\ve_2\approx \ve_3\approx 1$ the global minimum of
entropy is provided by the solution $\{12\}\vee\{23\}\vee\{34\}$. It has
the same meaning as above: among actions of comparable cost the
risk-minimizing agent chooses the one with the lowest cost (i.e.,
$p_2>0$, but $p_3=p_4=0$) for $\xi_1\leq E\leq\xi_2$. Now this
solution does not exist for $\xi_2\leq E\leq\xi_3$. In this
regime the agent takes the solution $\{23\}$, where $p_2>0$, $p_3>0$,
but $p_4=0$. For $\ve_2\approx \ve_3\ll 1$ the global minimum
is $\{14\}$, as expected. Now $p_2=p_3=p_4>0$. These two regimes are
similar to those for $n=3$. For $\ve_1\approx\ve_2\ll\ve_3\approx \ve_4$
the global minimum $\{14\}$ for a relatively low $E$ changes to
$\{13\}\vee\{34\}$ for a large $E$. Finally, when both $\ve_2$ and $
\ve_3$ are close to $0.5$, there is a sequence of transitions from one
solution to another, such that every solution becomes a global
minimum for a certain $E$; e.g., for $\ve_2=0.4$ and $\ve_3=0.5$ we
observe the following transitions upon increasing $E$: $\{14\}\to
\{12\}\to \{24\}\to \{23\}\to \{34\}$. 

\section{Summary and Discussion}
\no{summa}

An agent who wants to be adaptive in choosing between several actions in
a varying and complex environment cannot exclusively focus on the
least-cost action.  The agent should also explore actions with
non-minimal costs, and not only exploit the action with the minimal
cost. This is the known exploration-exploitation trade-off, which exists
in various forms and fields. We here worked out a set-up that allows us
to study the trade-off within decision-making. The agent faces several
actions, whose initial costs are known, but it is not known how they
will change given the time and actions of the agent.  Now the agent
needs to perform several actions so that if there is a regular
information about costs, then this information can be gathered.  With
which probabilities should he choose initial actions in such an agnostic
situation? 

We worked out one possibility, where the exploration goes via
risk-minimization---a heuristic rule that people frequently apply in
uncertain situations. Risk is a wide notion that appears in various
situations. In particular, both risk-minimization (aversion) and
risk-maximization are seen in experiments with people
gambling on uncertain monetary outcomes~\cite{baron}. 

%where the majority of human subjects are risk-averse if they have to
%choose between sufficiently large (vs.  small) monetary
%gains~\cite{baron}. In contrast, they are risk-seeking if subject to
%sure losses (costs). 

In our situation the treatment of risk is easier than usual
(cf.~\cite{haim,aumann,leshno}) because we compare agents that have the
same costs for their actions and decide to invest the same amount of
average utility into exploration, i.e., into not taking the least-cost
action {\it only}. We start with the stochastic dominance (or
majorization) condition, which constraints specific risk measures.
Given certain standard constraints, we show that the entropy can be
employed as a measure of risk. This result is obtained via several
different axiomatization schemes; see section \ref{rii}. Thus the
risk-averse (risk-seeking) agent will minimize (maximize) the entropy
given the average utility invested into adaptation, and also the
constraint that more costly actions should get less probability.
There are two different strategies of risk aversion: putting more
probability on low-cost vs. high-cost actions. These strategies
are different, and they relate to a rich behavior spectrum even in the
simplest case of three actions. 

While the entropy maximization is a well-known rule in 
probabilistic inference \cite{jaynes,shore,ttl,uffink,grun,landes1,landes2,balian,good,christensen},
the entropy minimization is an under-explored idea; cf.
\cite{lindsay,polgar,jaynes_pra,huhu,karen,good,christensen}. We show that
this method can lead to useful predictions, e.g., it recovers (under
definite conditions) the $\epsilon$-greedy probability known in
reinforcement learning theory \cite{barto}. Our main result is that the
entropy-minimizing agent (in contrast to the entropy-maximizing one)
shows certain aspects of intelligent behavior: {\it (i)} taking into
account costly actions; {\it (ii)} choosing the best alternative among
two comparable ones; {\it (iii)} cognitive dissonance. 

Within {\it (i)} the entropy-minimizing (risk-averse) agent puts more
probability into the least-cost action than the risk-seeking
(entropy-maximizing) agent, and distributes the remainder such that the
largest-cost action gets more probability than the risk-seeking
agent; see (\ref{a13}, \ref{b13}). Empirically, this scenario coincides
with the $\epsilon$-greedy strategy known in reinforcement learning
\cite{barto}, but now it is derived|together with its validity
conditions|from a more general principle of entropy minimization.
Overall, this scenario resembles the behavior of a scientist who
follows the incremental character of any science|hence devotes most of
his time to traditional, low-cost subjects but is still open-minded
enough to venture on alternatives that have higher costs of
implementation and recognition. 

When choosing between two actions with comparable cost, the
entropy-minimizing agent chooses to put a higher probability on the
lower-cost action; cf. {\it (ii)}. We find this to be a (rudimentary)
scientific attitude. The progress of science does relate to noting small
differences in experiments and/or in theory; e.g. successful scientists
are prone to see potential contradictions that are ignored by others.
Also, modern theories of physics (e.g.,  quantum mechanics) are based on
small experimental differences between their predictions and those of
classical physics. In contrast, the entropy-maximizing agent gives
approximately the same probability for actions with close costs. 

The entropy-minimizing agent also demonstrates cognitive dissonance: by
increasing the amount of utility $E$ invested into adaptation, this
agent tends to nullify probabilities of high-cost actions. This effect
relates to the fact that the minimization of entropy produces several
local minima that compete with each other (frustration). Within the
cognitive dissonance theory this effect is taken as a general sign of a
loosely defined cognitive consistency. We relate it with
risk-minimization.  
 
We mention that maximizing the entropy of probability paths was proposed
as a scheme for the emergence of intelligent behavior~\cite{wissner}.
Though the mathematical details of the original proposal are unclear
\cite{kappen}, the proposal was generalized to social collective systems
\cite{mann}, and formalized within the convex analysis~\cite{fry}.  We
stress however that within the set-up studied here, the entropy
maximization did not show features of intelligent behavior.  Further
work is needed to connect the presented research with the ones reported
in~\cite{wissner,mann,fry}. There are also proposals of employing
entropy minimization for improving the performance of machine learning
algorithms~\cite{kovach}. Future research may clarify relations of this
proposal with the presented results. 

Another open problem is how to modify/continue the presented theory for
applying it to problems of creativity modeling. One difference here is
that in creative task solving it is the action space that has to be
conceived and understood, while in various types of decision theories
the action space is fixed. As examples show,\footnote{\no{fisherman}
{\it Two friends approach a river and want to pass it. They ask
a fisherman who has a boat to help them. The fisherman has two
conditions: only one person can be in the boat; the boat should be
brought back from where it is taken.} People normally start solving this
task by over-concentrating on one specific (subjectively most likely)
option: The two friends together approach the same side of the river. This
would be a usual scenario for friends, but this makes the problem
unsolvable. Insisting on this option, people come up with rather
artificial constructions for solving the problem. But it is nowhere said
that the friends approached the same side of the river. If they 
approached the different sides, the problem has a trivial solution,
which would be found if people devoted some time to this (subjectively
less likely) possibility. } even with this serious difference there are
clear analogies between creative task solving and the
exploration-exploitation dilemma. 

\comment{ The minimization of entropy (\ref{entrop}) leaves several open
questions, in particular about the description of agents having a
different degree of risk-avoidance or risk-seeking. One wonders whether
this question can be addressed via taking a more general measure
of uncertainty, {\it viz.} the R\'enyi entropy; see Appendix \ref{reno} 
for preliminary considerations.}

\section*{Acknowledgment} A.E.A. thanks K.V. Hovhannisyan for useful
discussions. A.E.A. was supported by SCS of Armenia, grant 18RF-015. 

This research is based upon work supported in part by the Office of the
Director of National Intelligence (ODNI), Intelligence Advanced Research
Projects Activity (IARPA), via 2017-17071900005. The views and
conclusions contained herein are those of the authors and should not be
interpreted as necessarily representing the official policies, either
expressed or implied, of ODNI, IARPA, or the U.S. Government. The U.S.
Government is authorized to reproduce and distribute reprints for
governmental purposes notwithstanding any copyright annotation therein.

\clearpage

\clearpage

\appendix

\section{Alternative route to constrained entropy optimization}
\label{ko}

The purpose of this Appendix is to outline an alternative method for
obtaining entropy as a measure of risk and uncertainty; see
(\ref{lagri}). Recall that optimization (i.e., minimization or
maximization) of entropy (\ref{entrop}) under constraint (\ref{eq:4})
can be done via the optimization of the Lagrange function \be \no{lagr}
{\cal L}= -\sum_{k=1}^np_k\ln p_k+\hat\beta
\sum_{k=1}^np_k\varepsilon_k, \ee where $\hat\beta$ is the Lagrange
multiplier that corresponds with $E$ in (\ref{eq:4}).
Following~\cite{aczel} we shall mention an approach that allows us to
recover directly (\ref{lagr}) via few reasonable axioms.  This is useful
as an alternative (and more direct) route to optimizing entropy under
(\ref{eq:4}). 

Let us reinterpret actions $\A_1,...,\A_n$ in (\ref{eq:1}) as events of
a classical probability space. One seeks a measure ${\cal L}_n(\A_1,
p_1; ...; \A_n, p_n)$ of uncertainty (or risk) that depends on both the
probabilities and the corresponding events and that hold the following
axioms~\cite{aczel}: 

\begin{enumerate}[{(a)}]

\item\label{a}
${\cal L}_n(\A_1, p_1; ...; \A_n, p_n)$
is symmetric with respect to any permutation of $n$ elements
$(z_1,...,z_n)$, where $z_k=(\A_k, p_k)$.

\item\label{b}
${\cal L}_n(\A_1, p_1; ...; \A_n, p_n)$ holds the branching feature
\be
\no{branch}
{\cal L}_n(\A_1, p_1; ...; \A_n, p_n)={\cal L}_{n-1}(\A_1\cup \A_2, p_1+p_2;\A_3, p_3; ...; \A_n, p_n)
+(p_1+p_2){\cal L}_2(\A_1, \frac{p_1}{p_1+p_2}; \A_2, \frac{p_2}{p_1+p_2}).
\ee
This is a natural feature for an uncertainty, where joining to events
$\A_1$ and $\A_2$ (thus $p_1+p_2$ is the joint probability) leaves the
residual uncertainty ${\cal L}_2(\A_1, \frac{p_1}{p_1+p_2}; \A_2,
\frac{p_2}{p_1+p_2})$ with conditional probabilities
$\frac{p_1}{p_1+p_2}$ and $\frac{p_2}{p_1+p_2}$ for $\A_1$ and $\A_2$,
respectively. 

\item\label{c}
${\cal L}_n(\A_1, p_1; ...; \A_n, p_n)$ is a
continuous function of $p_1, p_2,...,p_n$. 

\end{enumerate}

The three axioms lead to~\cite{aczel}:
\be
{\cal L}_n(\A_1, p_1; ...; \A_n, p_n)= -\sum_{k=1}^np_k\ln p_k+\hat\gamma
\sum_{k=1}^np_k\varepsilon_k(\A_k),
\no{contr}
\ee
where $\varepsilon_k(\A_k)$ is an arbitrary function of $\A_k$, and
$\hat\gamma$ is an arbitrary constant. (An irrelevant additive constant
was fixed to zero). Interpreting $\varepsilon_k(\A_k)$ as the cost
related to $\A_k$, and equating $\hat\gamma=\hat\beta$ we revert from
(\ref{contr}) to (\ref{lagr}). Eq.~(\ref{contr}) and many related
results can be proved via the functional equations methods reviewed
in~\cite{ali}. 

Note, that expressions similar to (\ref{contr}), i.e., a convex
combination of entropy and expected cost (negative utility) were proposed
in~\cite{chin1,chin2} as a measure of risk. Refs.~\cite{chin1,chin2}
employ this measure for elucidating several controversies in the
decision theory. The same measure was axiomatically deduced and studied
in~\cite{ng}. Taking into account the axiomatic development, one can say
that the measure of risk (\ref{contr}) expressed by a linear combination
of entropy and expected cost does have normative features.

\section{Local minimality of solutions (\ref{eq:1000}).   }
\label{samson}

The aim of this Appendix is to show that the solutions for
entropy minimization given by (\ref{eq:1000}) do provide local minima of
entropy. This is an important point, because once the local minimality
is established, the solutions become meaningful even if they do
not provide the global entropy minimum. To illustrate ideas, we start 
with the simplest non-trivial situation. 

\subsection{$n=3$}

To check the local minimality we represent the probabilities of
different actions as [see (\ref{eq:8})]
\begin{eqnarray}
  \label{eq:26}
&&  p_1=\mu_1+\frac{\mu_2}{2}+\frac{\mu_3}{3}+
\lambda_1+\frac{\lambda_2}{2}+\frac{\lambda_3}{3}, \\
&&  p_2=\frac{\mu_2}{2}+\frac{\mu_3}{3}+
\frac{\lambda_2}{2}+\frac{\lambda_3}{3}, \\
\label{eq:266}
&&  p_3=\frac{\mu_3}{3}+\frac{\lambda_3}{3},\\
\label{drust}
&& p_1\geq p_2\geq p_3,   
\end{eqnarray}
where the unperturbed probabilities are [see (\ref{ush}, \ref{dat})] 
\begin{eqnarray}
  \label{eq:24}
  \lambda_1+\lambda_2+\lambda_3=1, \\
  \lambda_1\xi_1+\lambda_2\xi_2+\lambda_3\xi_3=E, 
  \label{eq:244}
\end{eqnarray}
and where small perturbations $\mu_k$ hold
\begin{eqnarray}
  \label{eq:250}
  \mu_1+\mu_2+\mu_3=0, \\
  \mu_1\xi_1+\mu_2\xi_2+\mu_3\xi_3=0. 
  \label{eq:2500}
\end{eqnarray}
We recall from (\ref{eq:11}) that
\begin{eqnarray}
  \label{kret}
  \xi_1\leq \xi_2 \leq \xi_3.
\end{eqnarray}

To check the local minimality of the first solution we take
$\lambda_2=0$ in (\ref{eq:26}--\ref{eq:266}). Now (\ref{drust})
demands [in addition to (\ref{eq:250}, \ref{eq:2500})]
\begin{eqnarray}
  \label{eq:27}
  \mu_2>0.
\end{eqnarray}
Now we have for the entropy changes due to the perturbation:
\begin{eqnarray}
  \label{eq:28}
  \Delta S&=&-\sum_{k=1}^3p_k\ln p_k +\sum_{k=1}^3p_k|_{\mu_l=0} \ln
  p_k|_{\mu_l=0} \\
\label{blin}
&=&-\sum_{k=1}^3(p_k-p_k|_{\mu_l=0}) \ln
  p_k|_{\mu_l=0} \\
  \label{eq:288}
&=& -(\mu_1+\frac{\mu_2}{2}+\frac{\mu_3}{3})
\ln\left[\frac{p_1|_{\mu_l=0}}{p_2|_{\mu_l=0}}\right]\\
  \label{eq:2888}
&=& \frac{2\mu_2}{3}\left[
  \frac{\xi_3-\xi_2}{\xi_3-\xi_1} -\frac{1}{4}
\right]\ln\left[\frac{p_1|_{\mu_l=0}}{p_2|_{\mu_l=0}}\right],
\end{eqnarray}
where in (\ref{blin}) we kept only the linear order over $\mu_l$, and
where we employed (\ref{eq:250}, \ref{eq:2500}) in (\ref{eq:288}) and in
(\ref{eq:2888}). Due to (\ref{eq:27}) and to
$\ln\left[\frac{p_1|_{\mu_l=0}}{p_2|_{\mu_l=0}}\right]\geq 0$, we get
that $\Delta S>0$ (we are looking for the local minimum of entropy) is
achieved for
\begin{eqnarray}
  \label{eq:30}
    \frac{\xi_3-\xi_2}{\xi_3-\xi_1} > \frac{1}{4}.
\end{eqnarray}
This inequality always holds, once one recalls the definition of $\xi_k$; see
(\ref{dat}, \ref{eq:08}).

To check the local minimality of the second solution we take
$\lambda_1=0$ in (\ref{eq:26}--\ref{eq:266}). Now (\ref{drust})
demands [in addition to (\ref{eq:250}, \ref{eq:2500})]
\begin{eqnarray}
  \label{eq:a27}
  \mu_1>0.
\end{eqnarray}
Repeating the same steps as in (\ref{eq:28}--\ref{eq:2888}), we get
\begin{eqnarray}
  \label{eq:29}
  \Delta S=
\frac{\mu_1}{3}\,
  \frac{\xi_2-\xi_1}{\xi_3-\xi_2}\,
\ln\left[\frac{p_2|_{\mu_l=0}}{p_3|_{\mu_l=0}}\right]>0,
\end{eqnarray}
i.e. this solution is local minimum (without additional conditions)
due to (\ref{eq:a27}),
$\ln\left[\frac{p_2|_{\mu_l=0}}{p_3|_{\mu_l=0}}\right]>0$ and
(\ref{kret}).  

For the third solution we take
$\lambda_3=0$ in (\ref{eq:26}--\ref{eq:266}). 
Now (\ref{drust})
demands [in addition to (\ref{eq:250}, \ref{eq:2500})]
\begin{eqnarray}
  \label{eq:b27}
  \mu_3>0.
\end{eqnarray}
We get
instead of (\ref{eq:29})
\begin{eqnarray}
  \label{eq:31}
  \Delta S={\cal O}(\mu_3)+
\frac{\mu_3}{3}\,
\ln\left[\frac{3\,e\,p_2|_{\mu_l=0}}{\mu_3}\right]>0.
\end{eqnarray}
This expression is always non-negative, whenever $\mu_3>0$ is
sufficiently small. Thus all solutions are always local minima.

\subsection{$n>3$}

We now turn to the more general situation and write probabilities as
\begin{eqnarray}
  \label{eq:c26}
&&  p_k=\sum_{l=k}^n \frac{\mu_l}{l} + \sum_{l=k}^n
\frac{\lambda_l}{l}, \qquad k=1,...,n,\\
\label{siser}
&& p_1\geq p_2\geq ...\geq  p_n, \\
\label{bor}
&& \sum_{l=k}^n \mu_l=0, \\
\label{meghu}
&& \sum_{l=k}^n\xi_l\mu_l=0,
\end{eqnarray}
where $\mu_l$ are perturbations. Now the unperturbed solution is
defined by only two non-zero elements in $\{\lambda_\alpha\}_{\alpha=1}^n\geq
0$: $\lambda_\alpha$ and $\lambda_\beta$. Eq.~(\ref{siser}) then
implies that besides $\mu_\alpha$ and $\mu_\beta$ all other $\mu_k$
are necessarily non-negative:
\begin{eqnarray}
  \label{eq:32}
  \mu_k\geq 0, \qquad k\not=\alpha, \qquad k\not=\beta.
\end{eqnarray}
Using (\ref{bor}, \ref{meghu}), $\mu_\alpha$ and $\mu_\beta$ are
expressed as
\begin{eqnarray}
  \label{eq:33}
\mu_\alpha=\sum_{k=1, k\not=\alpha, k\not=\beta}^n \mu_k
\frac{\xi_k-\xi_\beta}{\xi_\beta-\xi_\alpha},
\qquad
\mu_\beta=\sum_{k=1, k\not=\alpha, k\not=\beta}^n \mu_k
\frac{\xi_k-\xi_\alpha}{\xi_\alpha-\xi_\beta}.
\end{eqnarray}
Eqs.~(\ref{eq:33}) imply that if at least one $p_k|_{\mu_l=0}$ equals
to zero, the corresponding solution is locally stable via the same
mechanism as in (\ref{eq:31}).

Solutions for which $p_k|_{\mu_l=0}>0$ can be studied on the
case-by-case basis. For the solution with $\lambda_1>0$ and
$\lambda_n>0$ we obtain from (\ref{eq:33}) [cf. (\ref{eq:28})]: 
\begin{eqnarray}
  \label{sta}
  \Delta S&=&-\sum_{k=1}^n(p_k-p_k|_{\mu_l=0}) \ln
  p_k|_{\mu_l=0} \\
  \label{viator}
&=& \ln\left[\frac{p_1|_{\mu_l=0}}{p_2|_{\mu_l=0}}\right]\,
\frac{n-1}{n}\,\sum_{k=2}^{n-1}\mu_k 
\left\{
  \frac{\xi_n-\xi_k}{\xi_n-\xi_1} -\frac{n-k}{k(n-1)}
\right\}
\end{eqnarray}
Now using (\ref{dat}, \ref{eq:08}) for $\xi_k$ one can show
directly that all the curly brackets in (\ref{viator}) are non-negative,
which together with (\ref{eq:32}) and (\ref{siser}) implies $\Delta
S\geq 0$, i.e., this solution is a local minimum of entropy.
Generalizing this argument we converge to a conclusion that all the
solutions are local minima. 

\comment{
\section{The R\'enyi entropy}
\no{reno}

This entropy is defined as 
\be
\no{ren}
H_a=\frac{1}{1-a}\ln\left[\sum_{k=1}^np_k^a\right]\geq 0, 
\ee
with a positive parameter $a>0$~\cite{ren}. Now $H_a$ shares some (but not all)
features of the entropy; e.g. (\ref{ado}) holds, but (\ref{adon}) is not
valid.  At any rate, the maximization of the R\'enyi entropy is a valid
inference rule that generalizes the entropy maximization \cite{uffink}.
For $a\to 1$ we are back to the ordinary entropy (\ref{entrop}), while
for $a\to 0$ we get $H_a\to \ln n$ ($H_a$ is a decreasing function of
$a$). In contrast, for $a\to\infty$, we get $H_a\to -\ln p_{\rm max}$,
where $p_{\rm max}={\rm max}_k[p_k]$ is the maximal probability
(assuming for simplicity that it is unique). Admittedly, $ -\ln p_{\rm
max}$ still measures uncertainty, but in a rather weak sense; cf.
(\ref{major}) with $k=1$. Hence $a$ in the R\'enyi entropy can measure
the degree of risk-aversion or risk-seeking with the following limiting
cases. For $a\to 0$ both maximization and minimization of $H_a$ will
lead to similar results, since the agent who employs $H_{a\to 0}$ as a
measure of uncertainty will tend to see almost all probabilities as
maximally uncertain. For $a\to \infty$ the uncertainty will be measured
exclusively via the maximal probability $p_{\rm max}$, i.e. $p_{\rm
max}\to 1$ and $p_{\rm max}\to \frac{1}{n}$ mean (resp.) certain and
uncertain. 
}

\end{document}